\documentclass{aa}
\usepackage{graphicx}
\usepackage{txfonts}
\usepackage{ulem}
\usepackage{multirow}

\bibpunct{(}{)}{;}{a}{}{,} 

\begin{document}

\title{AGN in the ULIRG HE 0435-5304\thanks{based on observations made with the South African Large Telescope (SALT) and the Very Large Telescope (VLT)}}

\subtitle{}

\author{Krzysztof Hryniewicz\inst{1}, Małgorzata Bankowicz\inst{2}, Katarzyna Małek\inst{1,3}, Aleksander Herzig\inst{2}, Agnieszka Pollo\inst{1,2}
}

\institute{National Centre for Nuclear Research, Pasteura 7, 02-093 Warsaw, Poland\\
        \email{krzysztof.hryniewicz@ncbj.gov.pl }
        \and
        Astronomical Observatory, Jagiellonian University, Orla 171, 30-244 Krakow, Poland
        \and 
        Aix Marseille Univ, CNRS, CNES, LAM, Marseille, France
        }
        
\titlerunning{AGN ULIRG HE 0435-5304}
\authorrunning{Hryniewicz et al.}

\date{Received 30.03.2021; accepted 26.11.2021}

\abstract
{HE 0435-5304 from Hamburg European Southern Observatory survey (Hamburg-ESO, HE) is a quasar that appears in the literature with two conflicting redshift values: 
$\sim 1.2$ and $\sim 0.4$. It 
was used in the studies of the intergalactic medium through fitting of the narrow absorption lines in its ultraviolet (UV) spectrum. This source 
is also known historically as a luminous infrared galaxy (LIRG).}
{We present optical spectra of HE 0435-5304, aiming to precisely measure 
its redshift   and to study its physical properties. 
In particular, properties of its active nucleus, which is studied in the context of the source being identified here as an ultra-luminous infrared galaxy (ULIRG), 
allow us to place this quasar in the  context of the general population.
}
{We 
 analyzed optical spectra of the quasar HE 0435-5304. 
 Fitting the spectra, we focused on modeling 
H$\beta$ and [O III] lines. Based on these, we derived the virial black hole mass, bolometric luminosity, and Eddington ratio of the active galactic nucleus (AGN). Additionally, we performed 
broad band photometry fitting which allows us to quantify host galaxy parameters. 
Based on available mixed IR/optical/UV data spanning over a decade, we discuss the possible evolution of physical properties of the source and the influence of the observing conditions on our results.
}
{
The improved redshift value of HE 0435-5304  is estimated to $0.42788 \pm 0.00027$ based on the [O II] line --- the narrowest line in the spectra --- which is mostly consistent with the narrowest components of the other emission lines. 
The source was found to be a relatively 
massive and luminous AGN whose host galaxy is 
actively forming stars. Although its stellar population seems to be heavily obscured, we did not find evidence for significant obscuration of the
nucleus. We 
conclude that the AGN  HE 0435-5304  is a rather prominent iron emitter from the extreme type-A population very close to the narrow-line Seyfert 1 (NLSy1) group. The fact that the width of the H$\beta$ line appears to be systematically growing in its broadest component with time 
may suggest that this AGN is changing its broad line region (BLR). However, because of the influence of atmospheric effects contaminating spectral profiles, this finding is uncertain. 
}
{}


        \maketitle

\section{Introduction}

The quasar designated \object{HE 0435-5304}, which is located in the southern sky, is a bright object with a magnitude of 16.40 in the $V$ filter \citep{veron2010,souchay2015}. More recent observations taken by the Gaia Space Telescope provide a magnitude of 17.45 in the $G$ filter \citep{gaia2018}. This object is also luminous in ultraviolet \citep[UV;][]{seibert2012,monroe2016} and infrared \citep[IR; e.g.,][]{cutri2003}. Among others, it was observed as a part of the 
Akari Deep Field -- South survey by the Japan Aerospace eXploration Agency AKARI (meaning "light") satellite \citep[ADF--S; JAXA][]{shirahata2009}.

One interesting property of this quasar is the value of its redshift. Surveys which included a measurement of the redshift of HE 0435-5304 provide a value of $z=1.231$  \citep{wisotzki2000} or $z=0.425$ as estimated from the UV spectrum \citep{stocke2013}. The first observation was made as part of a catalog of 415 bright quasars and Seyfert-1 galaxies from the Hamburg European Southern Observatory (Hamburg/ESO) survey. 
The UV-based redshift value was obtained thanks to the Hubble Space Telescope (HST) Cosmic Origins Spectrograph instrument (COS). A value of $z=1.231$ is still given as the ``preferred redshift`` by the National Aeronautics and Space Administration Infrared Processing and Analysis Center Extragalactic Database (NASA/IPAC NED\footnote{https://ned.ipac.caltech.edu}). Both of the above values have been used in various analyses of this object. 
Circumgalactic medium studies by  \citet{stocke2013}, \citet{keeney2013}, \citet{danforth2016}, \citet{keeney2017}, \citet{keeney2018}, \citet{zheng2019}, \citet{richter2017}, \citet{fox2014},
and \citet{ribaudo2011} assume the value of $z=0.425$. 
At the same time, 
\citet{neelman2016} used the same HST COS UV spectrum  in a search for the damped Ly$\alpha$ absorbers 
assuming the redshift of $z=1.231$ measured by \citet{wisotzki2000}. As the interpretation of spectral lines used in the studies of intergalactic medium and quasars itself 
depends heavily on the proper redshift estimation,  an accurate value is vital.

Quasar HE 0435-5304 is also an interesting object in its own right. Being a very luminous far-infrared (FIR) source, 
it has been included in studies of star forming galaxies and  luminous  and ultra-luminous infrared galaxies (LIRGs and ULIRGs) conducted in the ADF--S by \citet{malek2014,malek2017}. However,  
the redshift value of $z=1.231$ used for the UV-to-IR spectral energy distribution (SED) fitting identified 
HE 0435-5304 as an extreme hyper-luminous infrared galaxy (HLIRG), with a dust luminosity of
$\log L_{\rm dust}=13.12\pm0.09 L_\odot $, and star formation and other dust properties strongly deviating from the usual trends for IR-bright galaxies. Considering the redshift value of $z=1.231$, this quasar would indeed be a very peculiar object. On the other hand, the best fitted photometric redshift for this object was found to be $z_{\rm phot}=0.3$ \citep{malek2017} 
With this value or with the value of $z=0.425$ given by \citet{stocke2013}, the best-fitting SED for HE 0435-5304 provides  a dust luminosity of $\log L_{\rm dust}=11.12\pm0.29 L_\odot$, which suggests that this object is an average LIRG, with star formation rate (SFR) and other physical properties following relations very similar to those of other galaxies of this class at similar redshift \citep{malek2017}.

 HE 0435-5304 is therefore a potentially interesting object with a quasar and star forming activity in the same galaxy.
Indeed, many active galaxies have a coexisting active nucleus and star forming activity \citep{tomczak2016,stanley2017,yang2018,suh2019,ding2020,kirkpatrick2020}, which could be due to events with an approximately simultaneous activation phase.Whether there exists a causal relationship between these two processes or they are rather induced by the same external cause remains an open question \citep{kormendy2013}.
One of the most natural suspects in the latter case are galaxy mergers. 
Mergers are known to naturally induce both nuclear activity and star formation in galaxies \citep{hernquist1989,hopkins2006}. In this scenario, LIRG and ULIRG would be episodes in the galaxy history occurring as consequence of a merger \citep{sanders1988}. After the merger, we expect that the binary super massive black hole (SMBH) exists for some period of time in the center of a newly formed galaxy before the two BHs also merge \citep{begelman1980,vanwassenhove2012}. The activity of a binary SMBH can be revealed by its radio morphology \citep{leahy1984,parma1985,schoenmakers2000,rodriguez2006}, X-ray morphology \citep{komossa2003,hudson2006}, optical emission line profiles \citep{liu2010}, jet precession \citep{camenzind1992}, or variation in its light curve \citep{sillanpaa1988,bon2016}.
However, the long- or mid-term optical variability can be attributed not only to the possible presence of a binary SMBH but also to precession of a misaligned accretion disk \citep{begelman1980}.
Although HE 0435-5304 does not reveal any peculiar radio morphology or double-peak emission line profiles, its optical light curve is too short to be used to look for long-term sinusoidal variation.

The present paper is constructed as follows. In Section 2 we present the data used in our analysis. Section 3 is devoted to presenting our methodology and adopted models. In Section 4 we show the outcome of our fitting procedure and the derived physical properties of the quasar. A 
discussion and conclusions are given in Section 5.
Throughout this paper we use WMAP7 cosmology \citep{Komatsu2011}: $\Omega_m = 0.272$,
$\Omega_{\Lambda}$ = 0.728, $H_0$ = 70.4 km s$^{-1}$ Mpc$^{-1}$. 

\section{Data}

\subsection{Broad-band multi-epoch photometry}

To construct a broadband SED and light curve we collect data from various surveys: AKARI, Wide-field Infrared Survey Explorer (WISE), DEep Near-Infrared Survey
of the Southern Sky (DENIS), Dark Energy Survey (DES), Catalina Real-time Transient Survey (CRTS), SkyMapper, and GAlaxy Evolution eXplorer (GALEX).

\begin{figure*}
        \centering
        \includegraphics[width=0.99\linewidth]{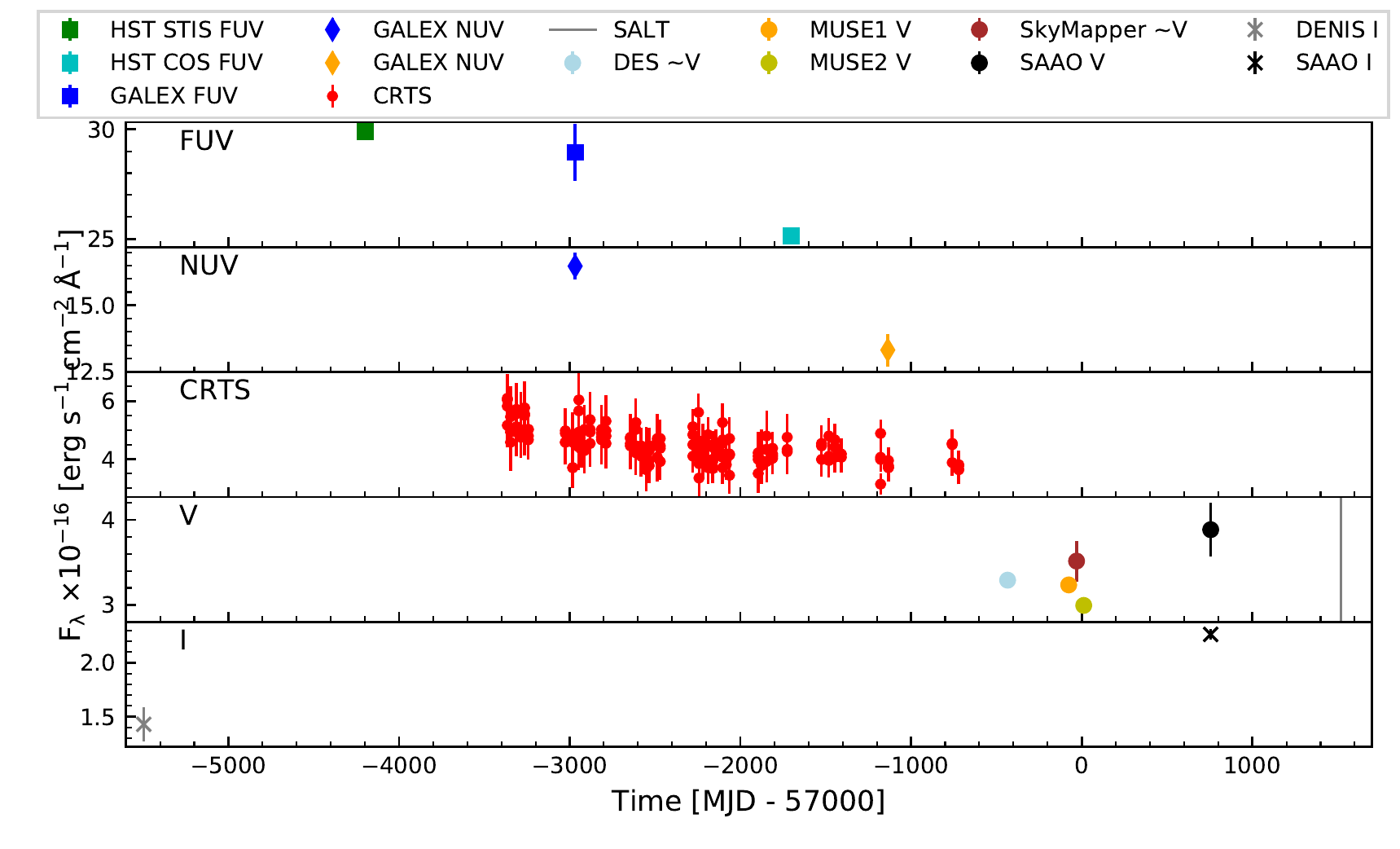}
        \caption{Light curve of HE 0435-5304 built from archival data. The different panels present light curves in the different photometric bands; from top to bottom: GALEX, HST FUV and NUV; CRTS clear (unfiltered) band; and Johnson V and I. We plot GALEX FUV photometry in blue with FUV derived from HST spectra STIS in green and COS in cyan. GALEX NUV observations in the second panel are plotted in blue and orange diamonds. CRTS points are presented in red. DES is plotted in navy blue and SkyMapper in brown which are interpolated to V wavelength using neighboring filters. Both MUSE spectra (in orange and yellow) allowed us to compute V points using filter transmission curves. SALT, calibrated using SAAO photometry, is plotted with gray stars as its spectrum only covers  this filter  entirely. SAAO in black were taken for V filter. For comparison of the epoch and flux, the IR photometry data of DENIS I (gray "x" symbols) and SAAO I (black "x" symbols) are plotted. }
        \label{fig:lightcurve}
\end{figure*}

\begin{itemize}
    \item 
The richest photometric dataset is provided by the Catalina Real-time Transient Survey \citep{drake2009}.
The photometric light curve provided by the CRTS  covers the period from 2005 to 2013. Those observations provide 
information on the time-variability of the source as registered by an unfiltered CCD. We present these data in Figure \ref{fig:lightcurve} with red points where we also mark other observations used in this article. There is one particular outlier in the Catalina light curve, a point with over two times higher flux. This is most probably an artifact, such as a cosmic ray, or (less likely) a sign of a single large intranight variability event, as other points gathered that night by CRTS are consistent with the overall light curve.
We gathered dates of the observations in Table~\ref{tab:dates} and present the light curve in Fig.~\ref{fig:lightcurve}.
    \item
AKARI, 
 originally ASTRO-F, is the Japanese infrared satellite operated from 2006 to 2011. It provided data in one all-sky survey and two deep surveys: North Ecliptic Pole (NEP) and ADF-S. AKARI observed the sky in the wavelength range from 9 to 180 $\mu m$. We use data from ADF-S deep field which provided data from four photometric bands centered at : 65, 90, 140, and 160 $\mu m$. The catalog in which HE 0435 is listed was released on 2008 October 17  and contains only WIDE-S measurements.
    \item
The DEep Near-Infrared Survey of the Southern Sky \citep{baudrand2007} operated from late 1995 through the end of 1999. HE 0435-5304 was observed in I filter on 1999 November 22.
    \item
The Wide-field Infrared Survey Explorer \citep{wright2010,mainzer2014} is also used. HE 0435-5304 was observed in the IR filters at 2010 January 23. We do not use WISE W4  because of the low accuracy of this filter.
    \item
Data from the Dark Energy Survey provided grizY photometry acquired on 2013 October 1 \citep{abbott2018}  add a point in the light curve a few weeks after the last CRTS point.  We increase uncertainty by adding statistical uncertainty on the shift for calibrator on HST, the single-epoch photometric statistical precision derived from repeated measurements of calibration stars and the median co-added zero-point statistical uncertainty mentioned by \cite{abbott2018}.
    \item
SkyMapper photometry in ugriz is also available \citep{wolf2018} for our target. The observation of HE 0435-5304 was taken on 2014 November 9, between two Very Large Telescope (VLT) spectra. The SkyMapper dataset fills the gap between points in optics and early GALEX observations in the wavelength regime with its "u" filter.
    \item
We use GAlaxy Evolution eXplorer \citep{martin2005} photometry resulting from two observational campaigns: AIS419 on 2006 October 24 with both far-UV (FUV) and near-UV (NUV) filters, and AIS420 conducted on 2011 October 30 in the NUV band. Derived fluxes were measured using the Kron radius of 3.5 arcsec.
\end{itemize}

We list photometric points together with the corresponding wavelength and observation dates in Table~\ref{tab:dates} and Figure~\ref{fig:lightcurve}. This set was collected to check the appearance of HE 0435-5304 and to decipher whether or not it is variable. It therefore contains only similar wavelength filters with multiple observations. The given wavelengths for photometric points are mostly taken from Spanish Virtual Observatory (SVO)\footnote{http://svo2.cab.inta-csic.es/theory/fps/} services (with the transmission curves). Values for points derived from spectroscopy were computed from transmission curves. We used archival HST spectra from the Space Telescope Imaging Spectrograph (STIS) and the COS to compute additional FUV points. Exceptionally, for the photometry derived from the South African Large Telescope (SALT) spectra, we  use  R filter as it was the only filter that fits within the observed spectral range. In the case of CRTS, a single point was exceptionally different at 56896 MJD with the flux around $1.0 \times 10^{-15}$ erg s$^{-1}$ cm$^{-2}$ \AA$^{-1}$. We consider this to be an artifact and ignored the outlier in Figure~\ref{fig:lightcurve}. If a given filter (like DES g and r or SkyMapper g and r) does not match the wavelength of the other filters, we use the neighboring filters from the same survey to compute the linear interpolation in the logarithmic space to derive generic Johnson V flux at its average wavelength and plot it in Figure~\ref{fig:lightcurve}. The errors for those points were simply the square root of the summed squared errors of the filters used. We separately list filters used to build the SED in Table~\ref{tab:photometric_data}.

When all the photometry is assembled together, three points are systematically higher in flux, namely: GALEX FUV, GALEX NUV, and the mean CRTS unfiltered point. This GALEX observation was carried out on 2006 October 24 while the CRTS point is an average around the same date. The majority of the optical coverage comes from DES and SkyMapper which carried out observations on 2013 October 1 and 2014 November 9, respectively. During 2006, the source was brighter than in 2013/2014. To prevent a potential systematic difference in photometry from introducing an artificial offset between GALEX and SkyMapper, we computed optical photometric points from Multi Unit Spectroscopic Explorer (MUSE) spectral cubes assuming the same aperture as in GALEX (Kron radius 3.5 arcsec). Roughly 1.5 month passed between SkyMapper and MUSE expositions. The computed MUSE points are consistent with the SkyMapper point, and therefore we assume no other systematic offset in the photometry apart from the physical difference in the source luminosity.

\begin{table*}[h]
\caption{Dates of the observations made with different instruments.}
\label{tab:dates}
        \centering
\begin{tabular}{lllllll}
\hline
 Date       & Facility   & Type   & Filters   &   $\lambda_{mean}$ & $F_{\lambda}$  & Reference     \\
 &            &        &           & [\AA]             & [$\times 10^{-16}$ erg s$^{-1}$ cm$^{-2}$ \text{\AA}$^{-1}$]  &  \\
\hline
 1999-11-22 & DENIS      & p      & I         &               7930 & 1.43 $\pm$ 0.16     &           (1) \\
 2003-06-11 & HST STIS   & s      & FUV       &               1545 & 29.89 $\pm$ 0.20    &            \\
 2005-09-21 & CRTS start & p      & clear     &               6178 & 4.89 $\pm$ 0.68     &           (2) \\
 2012-12-18 & CRTS end   & p      & clear     &               6178 & 4.42 $\pm$ 0.52     &           (2) \\
 2006-10-24 & GALEX      & p      & FUV       &               1545 & 28.9 $\pm$ 1.4      &           (3) \\
 2006-10-24 & GALEX      & p      & NUV       &               2345 & 16.48 $\pm$ 0.51    &           (3) \\
 2011-10-30 & GALEX      & p      & NUV       &               2345 & 13.32 $\pm$ 0.62    &           (3) \\
 2010-04-13 & HST COS    & s      & FUV       &               1545 & 25.151 $\pm$ 0.033  &            \\
 2013-10-01 & DES        & p      & g         &               4862 & 4.102 $\pm$ 0.035 &           (4) \\
 2013-10-01 & DES        & p      & r         &               6460 & 2.566 $\pm$ 0.019 &           (4) \\
 2013-10-01 & DES        & i      & V        &               5524 & 3.290 $\pm$ 0.040 &            \\
 2014-09-24 & MUSE       & s      & V         &               5524 & 3.4125 $\pm$ 0.0062     &            \\
 2014-12-21 & MUSE       & s      & V         &               5524 & 3.2494 $\pm$ 0.0055     &            \\
 2014-11-09 & SkyMapper  & p      & g         &               5099 & 3.80 $\pm$ 0.24     &           (5) \\
 2014-11-09 & SkyMapper  & p      & r         &               6157 & 3.152 $\pm$ 0.053   &           (5) \\
 2014-11-09 & SkyMapper  & i      & V        &               5524 & 3.52 $\pm$ 0.25     &            \\
 2017-01-03 & SAAO       & p      & V         &               5524 & 3.88 $\pm$ 0.32   &        (6)    \\
 2017-01-03 & SAAO       & p      & I         &               8859 & 2.263 $\pm$ 0.047   &   (6) \\
\hline
\end{tabular}

\tablefoot{`Type' column denotes if data product is photometry (p), was computed from spectroscopy (s), or was interpolated from neighboring filters (i). The subsequent columns present filter name, observed mean wavelength of the filter, and observed flux. In the last column, we give references to the original photometric measurements, catalogs, or surveys used to obtain the flux values.}
\tablebib{(1) \cite{baudrand2007}; (2) \cite{drake2009}; (3) \cite{martin2005}; (4) \cite{abbott2018}; (5) \cite{wolf2018}; (6) \cite{bankowicz2020}.
}
\end{table*}

To achieve quasar flux calibration in the SALT spectrum, BVRI photometry from South African Astronomical Observatory (SAAO) 1 m telescope was used. Observations were made during 4 days of observations from 2016 December 29 to 2016 December 31 and on 2017 January 3. Nights were clear with the Moon phase between 0.5\% on the first night and with 29.2\% during the last night, with seeing was 1.6 arcsec. For every day of observations, the Moon was below the horizon. We acquired 78 science images with 20 images for filters B, V, and R and 18 images in filter I, along with zero (BIAS), dark thermal current (DARK), and sky flat-field (FLAT) images. Reduction was made with the use of the ESO Munich Image Data Analysis System \citep[ESO-MIDAS,][]{esomidas,midas}. Photometric measurements were made with the use of Dominion Astrophysical Observatory PHOTometry software \citep[DAOPHOT,][]{daophot,stetson1987} included in the Python version of the Image Reduction and Analysis Facility and its Python wrapper \citep[IRAF, PyRAF,][]{iraf,iraf93,pyraf} software and Python SALT reduction package \citep[PySALT,][]{pysalt}. The full width at half maximum (FWHM) of the target object HE0435-0543 was measured for each image along with background luminosity for proper photometry. Uncertainties for the measured magnitudes were provided by the IRAF software according to the algorithm implemented in the DAOPHOT package. The magnitude value for each filter was estimated as a weighted mean from each image acquired in the respective filters. A photometric calibration was performed using an image of the star with known magnitude measured with the same instrument. Using the Gaussian method of error propagation, a single magnitude uncertainty value was calculated. The measurement of magnitude performed within IRAF was with an uncertainty magnitude of 0.001, which is only an estimation of the error for the magnitude measurement alone. As a main source of uncertainty, we took the standard deviation of the background luminosity in the given filter. This led to estimated uncertainties of order 0.01--0.1 in magnitudes. 
Corresponding flux values are given in Table~\ref{tab:dates}.

\subsection{Optical and UV spectra}

We use three optical spectra of HE 0435-5304. Two of them were taken on 2014 September 24 (hereafter MUSE1 or M1) and 2014 December 21 (hereafter MUSE2 or M2) with the MUSE instrument on the Very Large Telescope under proposal 094.A-0131(B) by Joop Schaye and are now publicly available. These were retrieved from the European Southern Observatory (ESO) archive\footnote{https://archive.eso.org/scienceportal/home}. The MUSE data were reduced and calibrated with the ESO pipeline \citep{weilbacher2020}. We did not improve on this reduction and calibration because our target is bright and centered in the field of view. To handle spectral cubes we use the MUSE Python Data Analysis Framework (MPDAF) library \citep[MPDAF][]{mpdaf}. The final spectra cover the wavelength range of 3328--6552~\AA\ in the object rest-frame. In order to obtain a single flux value corresponding to a given wavelength, we spatially integrated flux density within an aperture radius of 2.0 arcseconds. This gives the optimal spectra quality. Increasing the aperture to 3.5 arcseconds gives 5.5 \% higher total flux with stronger atmospheric emission residuals for M1.

A third spectrum was acquired with SALT using the Multi-Object Spectroscopy (MOS) mode on the Robert Stobie Spectrograph (RSS) under proposals 2018-2-DDT-001.20190204 and 2018-2-DDT-001.20190207. This mode requires the use of a pre-prepared mask with slits carved in the positions of target sources. We used 1.5 arcsec slits with grating with 900 lines per millimeter covering the wavelength range from 5611 to 8589~\AA. The observations were performed in two days in two different settings with the camera shifted so that the gaps between the chips of the camera were on different wavelengths. Averaging these two sets of images provides continuous spectra with no gaps present. The first two exposures taken over 1200 seconds on 2019 February 4 were made with the Moon phase of 0\% with 1.4 arcsec seeing and a clear sky. A second set of two exposures, 1200 seconds each, was taken on 2019 February 7 with the Moon phase of 9\% with 1.6 arcsec seeing with a clear sky for the first exposure and a slightly cloudy sky during the second exposure. The second exposure was taken with seeing >1.6 arcsec but with less than 2 arcsec overall. Moonset was before the observational window and moon rise was after this period. For wavelength calibration, observations of two different lamps for different days were used, Copper-Argon lamp for the 2019 February 4 and Argon lamp for 2019 February 7. Data was pre-reduced by SALT staff members. Fits image consisted of 20 aperture observations from MOS with 16 galaxies and 4 for mask alignment stars. Using IRAF software, we extracted single-aperture two-dimensional spectra of our target quasar HE 0435-5304 along with sky background. Further extraction of one-dimensional spectra applied background signal removal. For a one-dimensional spectrum, we applied a wavelength calibration, creating measurement-ready data.

We corrected the resultant spectrum for the PC04600 blocking filter transmission. 
Although the SALT spectrum was collected with a 1.5 arcsec slit with no absolute flux calibration, we used external photometry from SAAO described in the previous section to achieve flux normalization.
In the subsequent step, we applied the correction for atmospheric extinction. An average reddening curve was used that was constructed from measurements at the Sutherland site \citep{burgh2007}. We assumed 
airmass equal to 1.32.

\begin{figure*}
        \centering
        \includegraphics[width=0.99\linewidth]{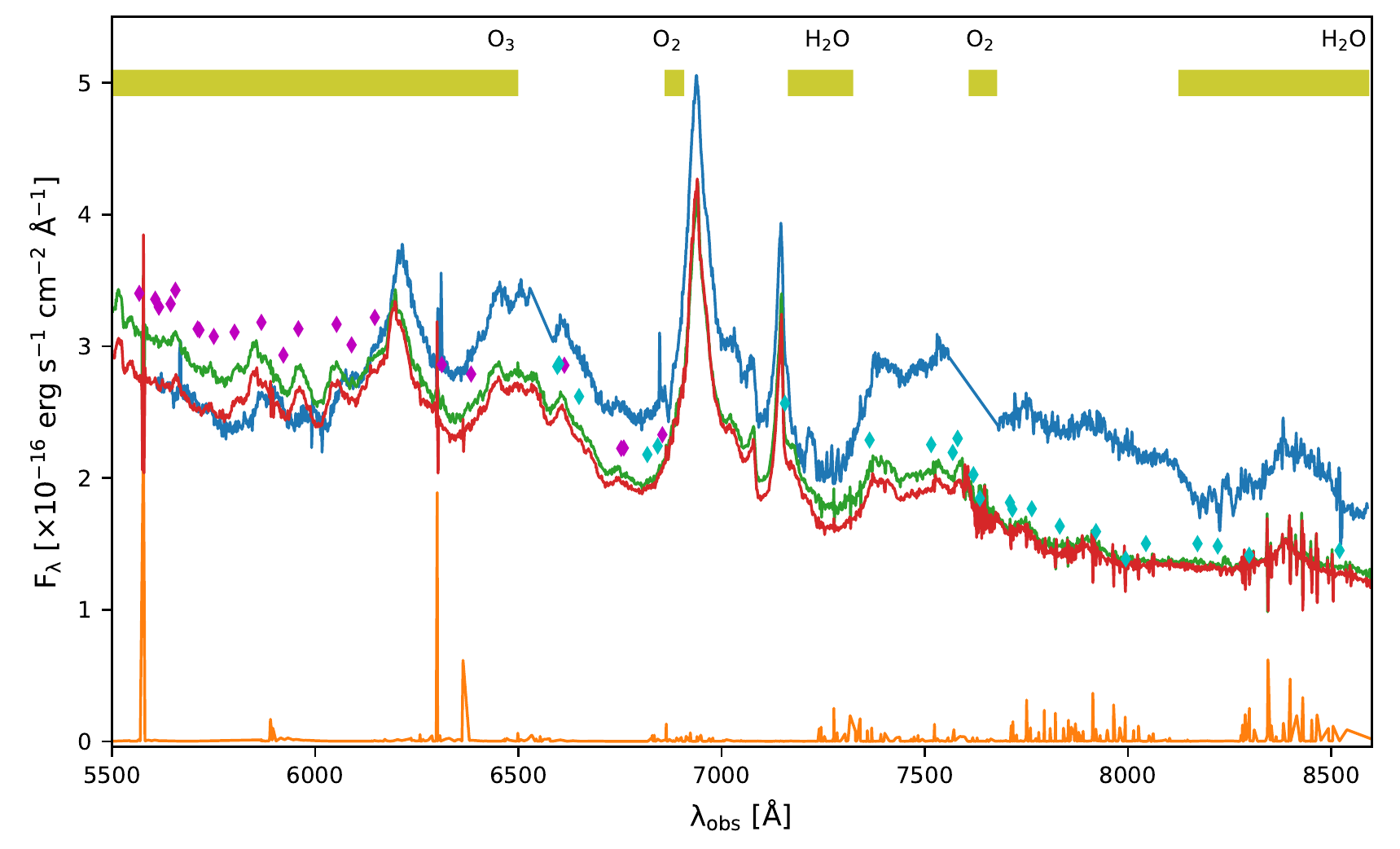}
        \caption{Comparison of the optical spectra of HE 0435-5304, atmospheric emission and IGM absorption. The MUSE spectra are presented as green and red lines, while the SALT spectrum is plotted as a blue line. The spectrum from SALT is rescaled for a convenient match with the MUSE data. Sky transmission spectrum is shown as an orange line. Hypothetical Balmer absorption lines from intergalactic medium at wavelengths corresponding to Lyman lines found by \cite{danforth2016} are marked by diamonds: cyan for H$\alpha$ and magenta for H$\beta$.}
        \label{fig:absorption}%
\end{figure*}

\begin{figure*}
        \centering
        \includegraphics[width=0.49\linewidth]{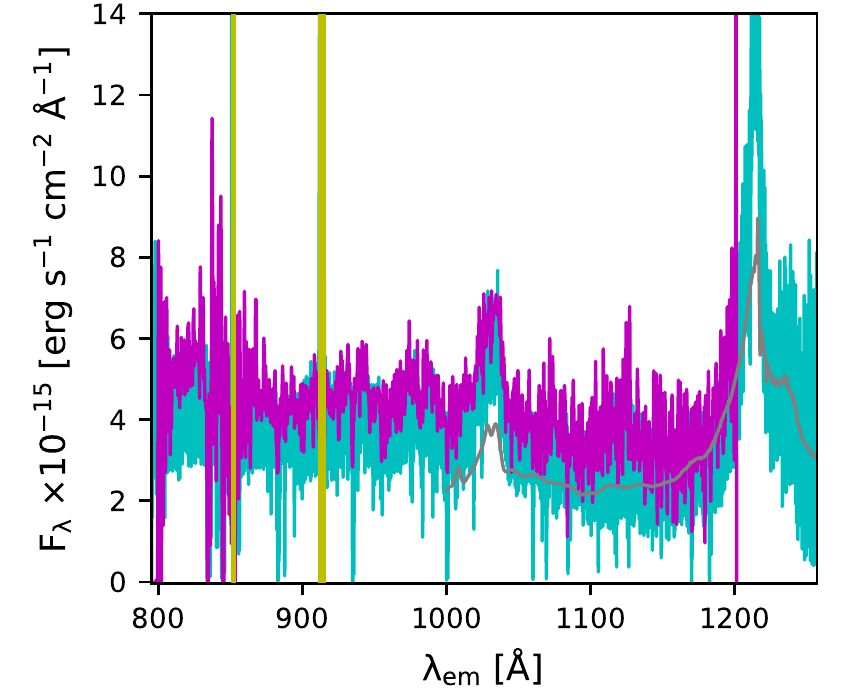}
        \includegraphics[width=0.49\linewidth]{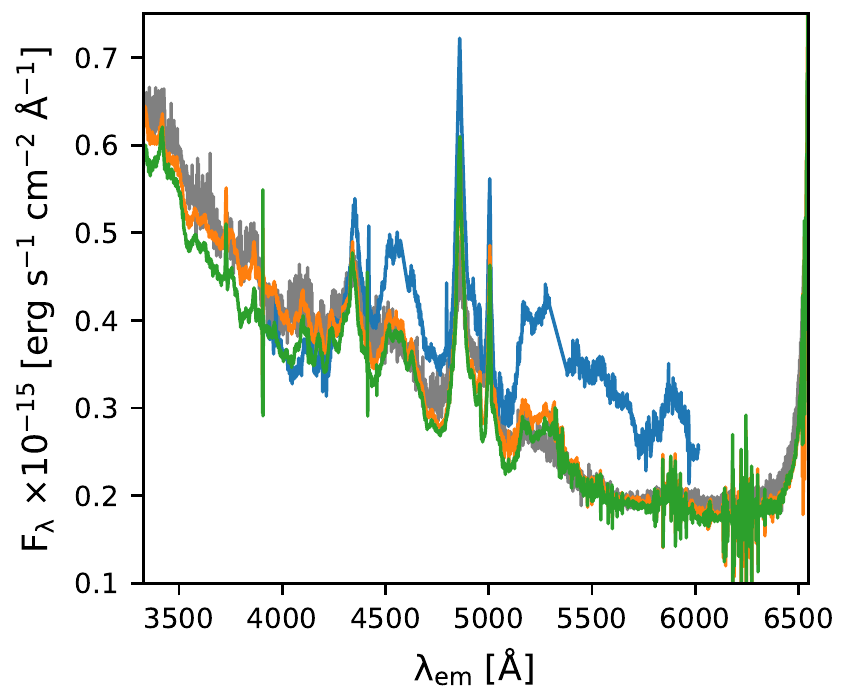}
        \caption{Composite spectra of HE 0435-5304. Left panel: HST STIS spectrum in magenta, HST COS data in cyan. The yellow region masks MW emission, and the composite by~\cite{selsing2016}  is shown in gray. Right panel: The spectrum from SALT is plotted in blue, MUSE spectra are presented in orange and green, and the composite is shown in gray.}
        \label{fig:spectra}%
\end{figure*}

We fitted a quadratic spline to both the uncorrected SALT spectrum and the SAAO photometry. The SALT spectrum was divided by its spline fit and multiplied by the SAAO photometry fitted spline. This is how we got rid of the strong distortion present due to the vignetting and other effects on the spectral shape. This also allowed us to achieve flux calibration matching the photometric level. 
However, the photometric observations were obtained over two years earlier than the spectroscopy, and therefore we renormalized the flux --- using the narrowest core component of the [O III] 5007 \AA\ line as measured in model B --- to the corresponding component of the M1 spectra model B fit (which is given in the results section). As those components have consistent kinematical shifts and widths, we assumed that this component should not change on the timescale of five years. This is roughly in line with the narrow line region reverberation study by \cite{Peterson2013}, where variation is detected on a timescale of the order of a decade.
The final corrected spectrum has an almost power law continuum shape and is presented in Fig.~\ref{fig:absorption} as the blue line. 
The MUSE spectra are visualized with green and red lines. To check for further problems, we compared our data with the sky transmission spectrum obtained with the VLT \citep{hanuschik2003}, which is shown by the orange line. Besides the obvious presence of sky residual features, we see two major artifacts at 6870~\AA\ and 7680~\AA. The first one is common to many SALT observations of different sources. As it is present on the H$\beta$ blue wing where a strong rise in the flux is present, and intrinsic absorption may be present as well, we decided to compare it to the SALT spectra of another quasar HE 0413-4031 \citep{zajacek2020}. In the spectrum of HE 0413-4031, a very similar feature in the same observed wavelength range is present and situated at the far red end of the emission line wing in approximately the same region as the spectral plateau. We therefore use this feature to examine the possible extent and shape of the absorption. The affected observed wavelengths span between 6860~\AA\ and 6914~\AA. The shape of the absorption was modeled with a spline and was subtracted from the SALT spectrum. The drop in flux around 8590 \AA\ happens at the very edge of the red detector, and we conclude that it is probably an instrumental artifact. We decided to mask it and  
exclude it from the fitting process.
A comparison of the optical spectra with the composite by \cite{selsing2016} is shown in Fig.~\ref{fig:spectra}. The MUSE spectra are plotted as green and orange lines and SALT data are presented as a blue line. 

As HE 0435-5304 is known for its intergalactic medium absorption lines which were fitted in the UV range, we checked for the presence of such lines in the optical range as well. To this end, we took into account absorption systems fitted by \cite{danforth2016}. Taking into account the Ly$\alpha$ absorption, we recomputed its position to mimic hypothetical absorption systems for H$\alpha$ and H$\beta$. The expected positions are marked in Fig.~\ref{fig:absorption} by cyan and magenta diamonds for H$\alpha$ and H$\beta$, respectively. Although at least some of these absorption lines are supposedly present, they are much weaker than Ly$\alpha$ and have mostly negligible effects. We therefore did not take into account those systems as their  only contribution is the small wrinkles in the spectra.

The biggest differences between MUSE and SALT spectra are seen 
redward of the [O III] emission line and blueward of the H$\gamma$ emission line. In addition to the difference in the continuum flux level and its spectral slope, possible absorption is present in the SALT spectrum in the rest-frame wavelength ranges 3930--4320~\AA\ and 5020--5160~\AA\ where the relative flux level is lower and the spectral shape is distorted.

Photometric measurements and two spectra are available for the UV part of the spectrum, both obtained by 
the HST. 
The HST STIS spectrum was taken 
on 2003 June 11, and the HST COS spectrum was taken on 2010 April 13. We use reduced spectra published by~\cite{danforth2016}.
Interestingly, both spectra have a similar flux level in spite of the almost seven years difference between the epochs. We present them in the left panel of Fig.~\ref{fig:spectra}, where the HST STIS spectrum is also presented along with the HST COS spectrum. We mask a very narrow emission of O I $\lambda$1305~\AA\ and Ly$\alpha$ $\lambda$1215~\AA\ in the observed frame, as it is of Milky Way (MW) origin. We compare the spectra with the QSO composite of \cite{selsing2016},  which has been normalized to match the HST COS flux level around the Ly$\alpha$ line in the rest frame. This allows for easy visual comparison of the UV and optical data sets.

\section{Methods}

\subsection{Spectral fitting}

Observed data were corrected for extinction using a formula by~\cite{fitzpatrick1999} (using standard $R_V = 3.1$) with $A_V$ = 0.016 mag based on maps by~\cite{schlafly2011}. As HE 0435-5304 is located close to the southern ecliptic pole, the Galactic extinction is very mild, as expected.
The correction for the MW extinction introduces a very minor change in the optical spectrum as at the wavelength 5100~\AA\ the flux change is about 1\%.

We account for multiple components in the fitting process: AGN continuum, 
Fe II emission blend, hydrogen Balmer, oxygen, and neon emission lines.
At this stage, we assume that starlight from the host galaxy is 
negligible in the optical part of the spectrum of this source. 
This assumption is then confirmed by the results of the SED fitting described in the 
following section.

The AGN continuum was modeled with a simple power law as single-epoch spectral data cover a limited range of wavelengths; this is a reasonable approximation for the accretion disk continuum of an active nucleus in the optical range. The exact parametrization we adopted is
\begin{equation}\label{key}
F_{\lambda} = F_{5100\text{\AA}} \left( \frac{\lambda}{5100\text{\AA}} \right)^{\alpha_\lambda}
.\end{equation}

In the preliminary tests
to account for the Fe II emission, we initially consider a template chosen from empirical templates  by~\cite{boroson1992,veron2004,tsuzuki2006,dong2010} and theoretical templates by~\cite{bruhweiler2008}. We test several iron emission models because the most widely used empirical templates \cite{boroson1992,veron2004} rely on a very specific source NLSy1 type: I Zw 1. In some cases, such as that described in~\cite{hryniewicz2014}, a better fit to the type A quasar spectrum was achieved with the theoretical Fe II emission template by~\cite{bruhweiler2008}. 
Another reason for testing several models is that the spectral window of the MUSE spectra is wider than the templates used by \cite{boroson1992,veron2004}, and therefore for the widest spectral window we have to rely on the templates with the broadest wavelength coverage (mostly \cite{bruhweiler2008,dong2010}).
Each template was convolved with a Gaussian of appropriate FWHM to account for Fe II kinematical broadening. This was done with a grid of broadenings ranging from 900 to 9000 km s$^{-1}$ in steps of 200 km s$^{-1}$ . The number of the grid point was a free parameter in the fitting process. Each template was allowed to have a shift in the range -1000 to 1000 km s$^{-1}$.

It is clearly visible in the spectra that the prominent iron emission has a strong blend with sharper peaks on top of it. Therefore, we fit the iron emission as a two-component model with narrow lines of FWHM < 1500 km s$^{-1}$ and broader lines of FWHM > 1500 km s$^{-1}$ \citep[see][for other cases of a two-component Fe II]{dong2010}. We 
initially investigated different templates to achieve an optimal fit as the Fe II emission may vary between sources, and the templates themselves 
differ from one to another. A particular difficulty stems from the different wavelength coverage of Fe II models, which we address by comparing fit results in different spectral windows. The range 4200--5500~\AA\ was used to address the limited wavelength span in templates made by~\cite{boroson1992,tsuzuki2006}. In this initial run, we used all available templates. 
To consistently use different templates, we treated all of them in the same way, interpolated to the same wavelength grid, and convolved with Gaussians of a defined width, thus achieving a broadened template.
In this initial step, we analysed the fit quality with the different templates. After comparing possible solutions, we identified the template that shows the best performance and confirmed this choice by visual inspection, evaluating the agreement with the observed narrow details, the overall residuals, and accuracy in the 4500--4800~\AA\ range. However, the original broadening grid as distributed by~\cite{boroson1992} shows a better performance than the templates broadened using our procedure (especially in the 900--2000 km s$^{-1}$ range). We were therefore forced to consider a single-template approach. Also, the fact that the majority of works on the AGNs' population, virial mass formulas, etc., were derived with the use of ~\cite{boroson1992} template, convince us to apply those relations consistently with observables derived particularly with this Fe II template.

The final run in the 4200--5500~\AA\ range was performed using exclusively the chosen template, which was the most commonly used optical template by~\cite{boroson1992}. Here we use the original template grid with accurate broadenings computed by these latter authors in steps of 250 km s$^{-1}$. This allows us to present a fit that is directly comparable to the other sources and results presented in the literature. This case is denoted `model B'. Additionally, using a popular template means that a comparison with literature results for the population is meaningful.

Separately, to address the highest number of emission lines in the whole spectral range, we extended the fitting wavelength range to the broader window of (3500, 6600)~\AA. In this way, we obtained crude fits to the line parameters at wavelengths below 4200~\AA. 
Only Fe II templates covering a broader wavelength range were used in this case ---  the simulated template collection by \cite{bruhweiler2008}. This option is referred to here as model W.

We fitted permitted hydrogen Balmer lines: H$\beta$ $\lambda$4861.3~\AA, H$\gamma$ $\lambda$4340.4~\AA, and H$\delta$ $\lambda$4102.89~\AA\ 
and forbidden lines: [O III] $\lambda$5006.84~\AA, [O II] $\lambda$3727.3~\AA, [Ne III] $\lambda$3869~\AA, and [Ne V] $\lambda$3425~\AA. The strongest spectral lines, H$\beta$ and H$\gamma,$ were fitted with four Gaussian components as this is the smallest number that reasonably describes the H$\beta$ line shape. The [O III] $\lambda$5006.84 \AA\ line was fitted with two Gaussians as this was sufficient. The [O III] $\lambda$4959 \AA\ line was assumed to have the same profile as the fitted [O III] $\lambda$5006.84 \AA\ but with an amplitude  scaled down by 2.98 \citep{storey2000}.
The remainder of the lines were fitted using a single component. This decision was motivated by the poorer S/N of the lower flux lines and the fact that forbidden lines are narrower. This choice restricts the total number of free parameters, which is important to limit the overall demand in computational resources. The [O II] line, which is a doublet, $\lambda\lambda$3725,3727.3 \AA,\ but is by far the narrowest and the most symmetrical, was therefore fitted as a single component profile as well.
H$\beta$, H$\gamma,$ and [O III] lines were fitted most accurately in a narrower spectral range fit (especially model B) and those results are given unless otherwise stated, as those are the most accurate values. In the broader spectral range fit (model W) we mostly supplemented fitting results with H$\delta$, [O II], [Ne III], and [Ne V] profiles, although these are less accurate because of the poorer iron fit.

All components were fitted at once using static and dynamic nesting variants of Monte Carlo Markov Chains. We assumed no a priori preference on the parameter values, and so used uniform priors as the initial distributions. The result of the sampling process is a posterior probability distribution, from which we derive a best fit value and its uncertainty at 1$\sigma$ confidence level (distance to 0.1587 and 0.8413 quantiles). Considering uncertainties from wavelength calibration and the influence of the atmosphere, which we think is of the order of 1 \AA\ at the position of H$\beta$ ---which is further discussed in the section about redshift estimation---, corresponds to a position and shift accuracy of about 62 km s$^{-1}$. We do not add this systematic error to the values given for the fitting parameters in the text.

\subsection{SED fitting}

To investigate the importance of starlight in the optical part of the spectrum and to  independently estimate the main physical properties of the host galaxy and AGN, we performed broadband SED fitting with Code Investigating GALaxy Emission\footnote{https://cigale.lam.fr/}  \citep[CIGALE,][]{Burgarella:2005,cigale}. This was also done by~\cite{malek2017} as part of the ADF-S ULIRGs sample study; however in the present work we update the treatment of the AGN part.  We use a new branch of the CIGALE code, named X-CIGALE \citep{Yang:2020}, which includes the SKIRTOR AGN model \citep{Stalevski:2016}, complementing AGN models of \cite{Fritz:2006}. Contrary to the recipe used by \citealp{Fritz:2006}, this module takes into account the clumpy torus that surrounds the central engine; it also assumes a non-negligible amount of extinction for some type I~AGNs, and is therefore more suitable for modeling of broad line quasars like HE~0435-5305.

We use the delayed star formation history with an additional burst to model the young stellar population following \cite{Malek2018} who performed SED fitting of millions of galaxies detected in IR as part of the Herschel Extragalactic Legacy Project (HELP). We adopt a \cite{Chabrier:2003} initial mass function with solar metallicity and \cite{bruzual2003} stellar population models. We use a modified \cite{CF2000} attenuation law (with grayer slopes) as this was found to work better with IR-bright galaxies \citep{Malek2018}. Moreover, \cite{Buat:2018} found that the power-law model based on a modification of the recipe of \cite{CF2000}  gives very good agreement with the radiative transfer models. 

Dust emission was fitted with the \cite{Dale2014} model in order to reduce the number of free parameters (for this purpose we set the AGN fraction to zero and use only an $\alpha$ power-law slope in the relation between the dust mass $M_{dust}$ and radiation field intensity $U$). This module simplifies the SED fitting procedure, reducing the number of parameters, which are difficult to constrain without using mid-IR (MIR) and FIR measurements. Photometric data used for the SED fitting  of HE~0435-5305 are presented in Table~\ref{tab:photometric_data}.
We illustrate the data points of the SED and the fitted model in Fig.~\ref{fig:cigale}.

\begin{table}[h] 
\caption{Summary of the photometric data for HE~0435--5304 used for the SED fitting.}
\label{tab:photometric_data}
\centering
    \begin{tabular}{l l r c c}
      \hline
    Redshift  &  \multicolumn{3}{l}{z$_{spec} = 0.427$ }\\
     \hline
     Telescope/ & \multirow{2}{*}{Filter} & $\lambda_{mean}$ &  $S_{\nu}$    & $S_{err\mbox{ }\nu}$\\
      Instrument&        & ($\mu$m)  &   ($\mu$Jy)     & ($\mu$Jy)      \\ \hline
      GALEX & FUV & 0.1545 & 0.2320  &    0.0264\\
            & NUV & 0.2345 & 0.2950  &    0.0308 \\
     CRTS & Catalina V & 0.6178  & 0.4370  &   0.0570  \\      
     CTIO & DECam g & 0.4862 &  0.3277 &  0.0003 \\
          & DECam r & 0.6461 & 0.3640 &   0.0004 \\
          & DECam i & 0.7851 & 0.3473 & 0.0005\\
          & DECam z & 0.9199 & 0.4533 & 0.0009\\
          & DECam  Y & 0.9907 & 0.3609 & 0.0030\\
 SkyMapper & u & 0.3497 & 0.2500 &   0.0079\\
              & g & 0.5099 & 0.3299 &   0.0203\\
              & r & 0.6157 &  0.4003 &   0.0067 \\
              & i & 0.7778 &  0.4234 &  0.0098 \\
              & z & 0.9162 & 0.4341 &   0.0040\\ 
    DENIS  & I & 0.7930 & 0.3823 &  0.0423 \\
    2MASS & J & 1.2350 & 0.5423 &  0.0510 \\
           & H & 1.6620 & 0.6521 &  0.0716 \\
           & Ks & 2.1590 & 0.9860 &  0.0900 \\
    WISE & W1 & 3.3526 &  1.9590 &  0.0415 \\
         & W2 & 4.6028 &  3.3530 &  0.0618 \\
         & W3 & 11.5608 & 11.1100 &  0.1739 \\
    AKARI & Wide-S & 89.2033 & 33.8400 & 5.3300 \\
\hline
    
\end{tabular}
\tablefoot{ $S_{\nu}$ is the flux and $S_{err\mbox{ }\nu}$ is the error in ($\mu$Jy). $\lambda_{mean}$ is the center of the specific filter band given in $\mu$m.}
\end{table}

\section{Results}

\subsection{Redshift}

Using the most prominent lines with a very narrow component we can potentially estimate the redshift with high precision. Here we present the results of our estimations based on the M1 spectrum, as this was obtained with the best weather conditions, the lowest humidity and air mass, and the longest exposure.
A redshift value of  $0.427581 \pm 0.000048$ was derived from the narrowest component of the [O III] $\lambda$ 5006.843 \AA\ emission line  whose center is fitted in the observed spectrum at $7147.674 \pm 0.057$ \AA, additionally taking into account the MUSE wavelength accuracy of 0.2 \AA\ \citep{weilbacher2020}.
While using the [O II] $\lambda$ 3727.3 \AA\ line, which apparently peaks at 5322.12 \AA\ in the observed frame, the estimated redshift is $0.42788 \pm 0.00027$. Similarly, using the narrowest H$\beta$ $\lambda$ 4861.3 \AA\ and H$\gamma$ $\lambda$ 4340.4 \AA,\ we obtained corresponding values of $0.42781 \pm 0.00021$ and $0.42781 \pm 0.00024$. Other lines are of lower amplitude, which results in higher noise over the spectral profiles and therefore difficulty in visual verification, and so we do not use other lines in this step. For hydrogen lines, we  arbitrarily assume 1 \AA\ accuracy because of the complicated multi-component nature of the line with possible stellar absorption (although assumed negligible in the fitting process). Similarly, [O II], because of its doublet nature and lack of good Fe II template coverage, was used with higher wavelength error. As [O II] is the narrowest line in the spectrum and does not show strong asymmetry, despite it being affected by uncertain underlying Fe II, we consider it the best estimated because it is more certainly associated with the host galaxy. The remaining estimates are mostly consistent within the error bars. Therefore, our preferred redshift is $0.42788 \pm 0.00027$ and this value is used in the spectral fitting.

\begin{figure*}
        \centering
        \includegraphics[width=0.99\linewidth]{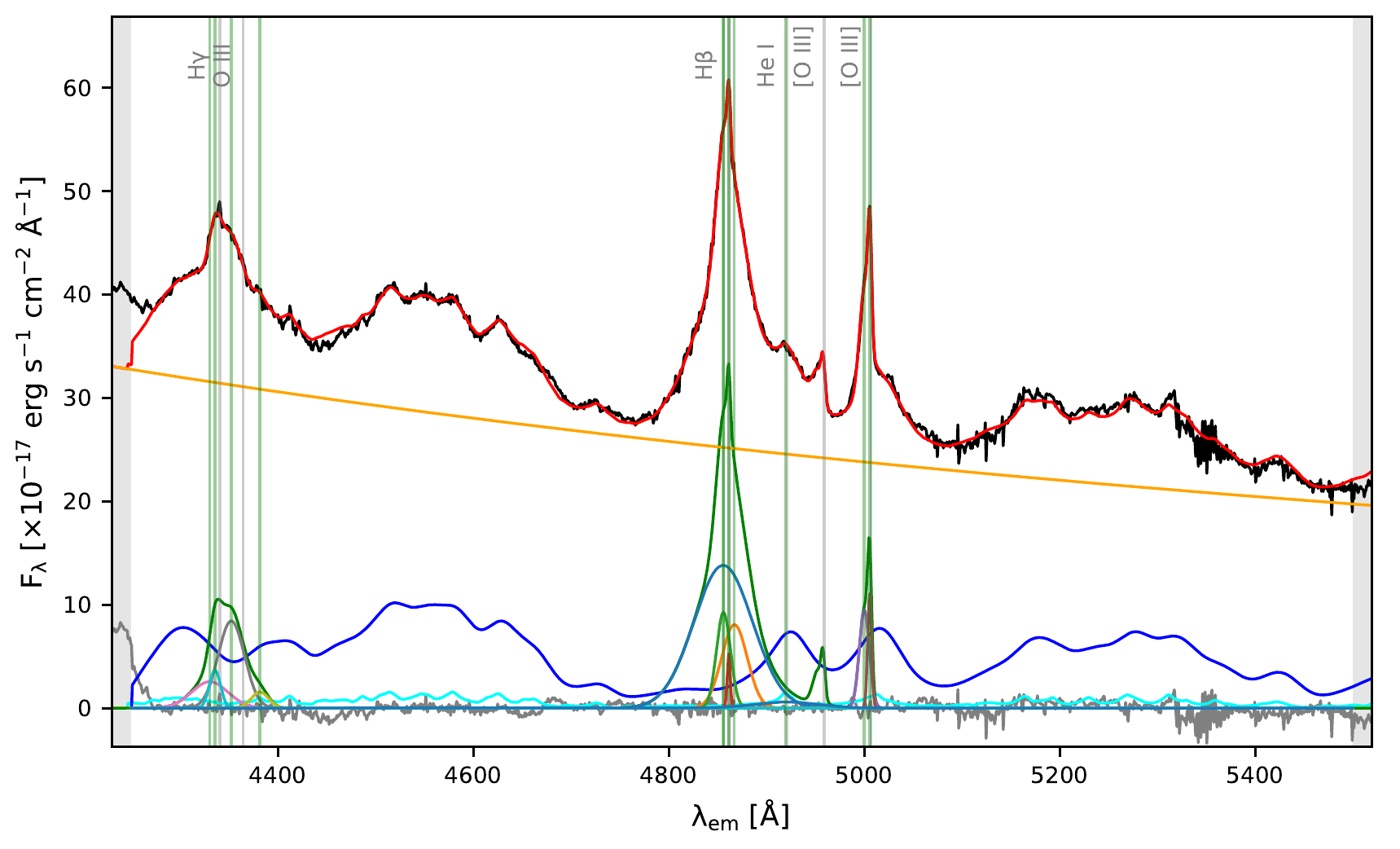}
        \caption{Fit to the M1 spectrum (model B), continuum fitted as a simple power law (orange) with the two components of blended Fe II emission (cyan and blue). The spectrum is plotted in black while the sum of components is presented with a red line. Overall emission line spectral profile fits are plotted in green while individual Gaussian components are presented in a range of other colors. The gray line presents the residuals. Vertical straight lines mark the expected laboratory wavelengths in gray and the fitted wavelengths in the separate components are shown in green.} 
        \label{fig:fitBM1}%
\end{figure*}

\begin{figure*}
        \centering
        \includegraphics[width=0.99\linewidth]{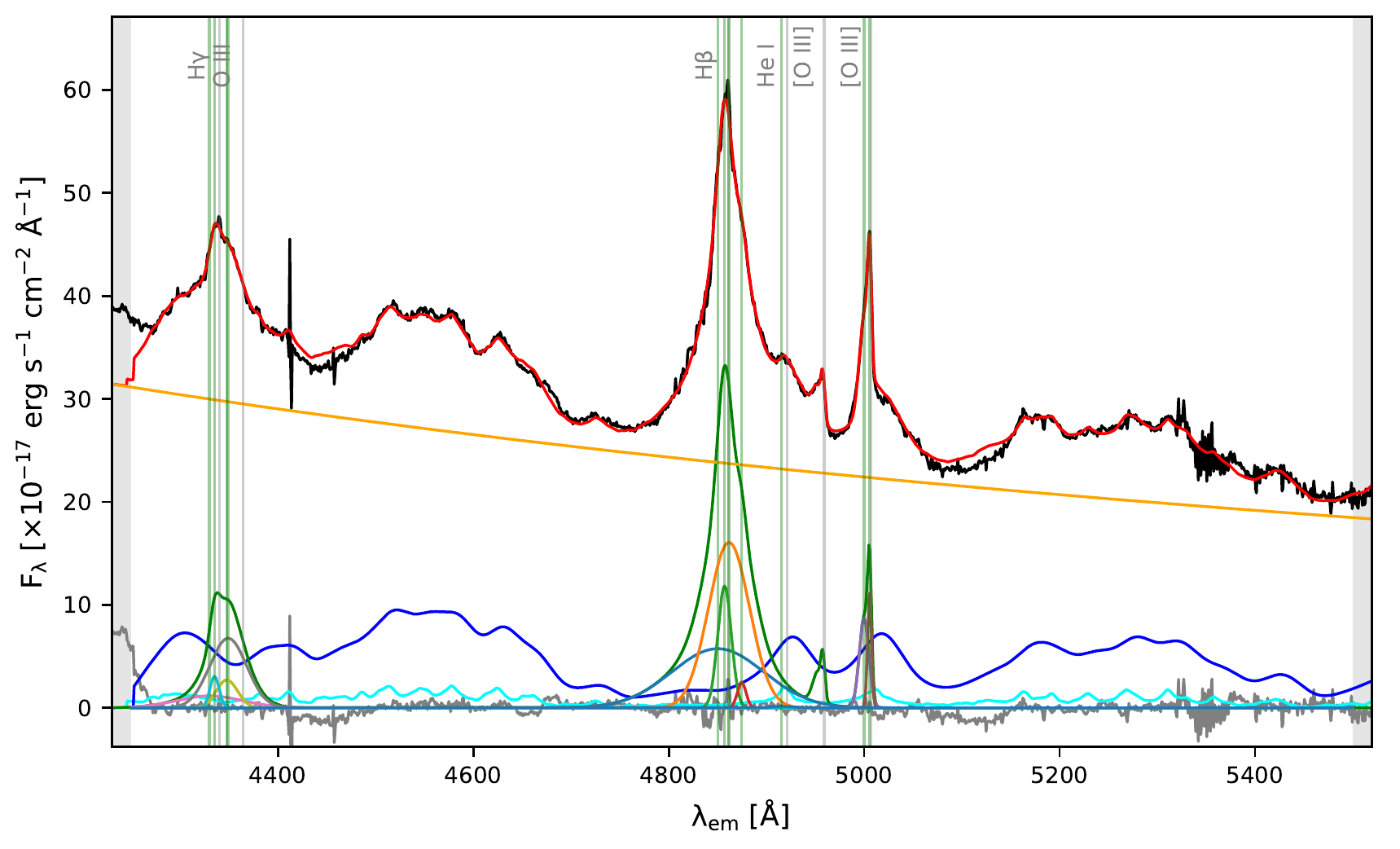}
        \caption{As in Fig.~\ref{fig:fitBM1} but fit to the M2 spectrum (model B).} 
        \label{fig:fitBM2}
\end{figure*}

\begin{figure*}
        \centering
        \includegraphics[width=0.99\linewidth]{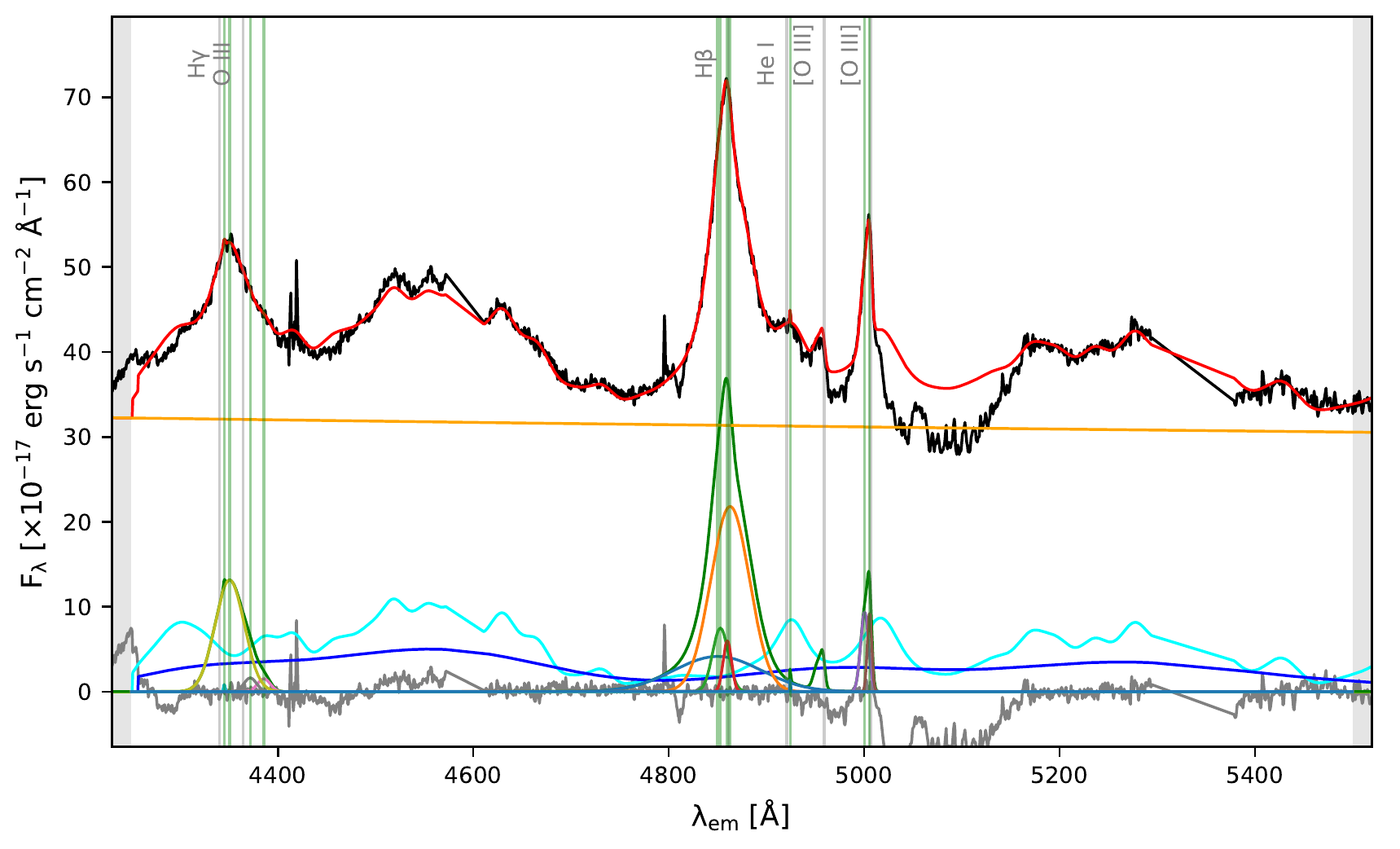}
        \caption{As in Fig. 4 but fit to the SALT spectrum (model B). } 
        \label{fig:fitBS}
\end{figure*}

\subsection{AGN}

Here we describe results of the fitting procedure, the aim of which is to estimate the emission line profiles with the multi-Gaussian model. Permitted lines are prominent and broad, which is generally characteristic of the emission from AGN. Overall, the continuum is blue and well approximated with a power law which is also characteristic of dominance from AGN emission in the UV and optical bands.

\subsubsection{Model B}

Among the Fe II templates  used here, the best fit was achieved with \cite{boroson1992} iron emission. 
The solution for M1 spectrum is presented in Fig.~\ref{fig:fitBM1} (named model B). This fit succeeds in highlighting the finer details of the iron emission. We therefore consider it to be our most accurate attempt to describe the data, and this is supported by the flat residuals.
We used a power law to fit the pseudo continuum. In the case of M1, the power $\alpha_{\lambda}$ = $-1.9620 \pm 0.0040$ and amplitude $F_{5100\text{\AA}} = (1.6030 \pm 0.0010) \times 10^{-16}$ erg s$^{-1}$ cm$^{-2}$ \AA$^{-1}$. These values indicate a rather blue spectrum but not as steep as in the composite by~\cite{selsing2016} as plotted in Fig.~\ref{fig:spectra}.

The amplitude at 5100~\AA~ gives a monochromatic luminosity of L$_{5100\text{\AA}}$ = $1.515 \pm   0.064 \times 10^{41}$ erg s$^{-1}$ \AA$^{-1}$. The uncertainty here accounts for both data and continuum fit uncertainties. On top of the continuum, two components of the strong Fe II blend are fitted. The shift of the templates is rather consistent with values of $-289 \pm 19$ km s$^{-1}$ and $62.8 \pm 7.3$ km s$^{-1}$ for the narrow and broad components, respectively. The narrow Fe II broadening was 900 $\pm$ 250 km s$^{-1}$ while the broad component was $2250 \pm 250$ km s$^{-1}$. The broad Fe II was dominating the total iron flux.
The fit parameter values are given in Table~\ref{tab:modelBM1}.
The two components of the Fe II model seem to be sufficient given the blended character of the complex iron emission, which which prevents us from deciphering whether or not there is a higher number of components present in the data.

In the M2 spectrum, taken 3 months after the M1 spectrum, the obtained
value of the continuum power index  was $\alpha_{\lambda}$ = $-2.0260 \pm 0.0036,$ with an amplitude of $F_{5100\text{\AA}} = (1.5095 \pm 0.0014) \times 10^{-16}$ erg s$^{-1}$ cm$^{-2}$ \AA$^{-1}$, thus reaching lower values than in the case of M1.
The iron template parameters were the same as in M1 in terms of widths: 900 $\pm$ 250 and $2250 \pm 250$ km s$^{-1}$, while the shift values were different: -344 $\pm$ 13 and 197.3$_{-7.9}^{+3.8}$ km s$^{-1}$.
In this case, the Fe II fit still accurately describes observations in the 4500-4700 \AA\ range, but the regions around 4450 \AA\ and 5100 \AA\ are much less well described than in the case of the M1 spectrum (see Fig.~\ref{fig:fitBM2}). For this reason, the realistic uncertainty on the continuum amplitude is much larger. 

Fitted individual emission lines are presented in Figure~\ref{fig:fitlines} where we show: H$\beta$ $\lambda$4861.3~\AA, H$\gamma$ $\lambda$4340.4~\AA, and [O III] $\lambda$5006.84 \AA\
lines fitted to the M1 spectrum.
 We provide measured flux and equivalent widths ($W_{\lambda}$) in Table~\ref{tab:modelBM1}.
The H$\beta$ spectral profile is very well described by the four Gaussian components. Its FWHM takes values of 157.0 $\pm$ 8.9, 722$_{-30}^{+33}$,     1397$_{-68}^{+59}$ , and     3128$_{-33}^{+37}$km s$^{-1}$ km s$^{-1}$ with shifts of 
-2.4$_{-5.4}^{+4.4}$,     -337 $\pm$ 16,      309$_{-58}^{+75}$, and     -338$_{-27}^{+26}$ km s$^{-1}$
km s$^{-1}$, respectively. The $W_{\lambda}$ values are 0.817$_{-0.061}^{+0.064}$,     6.24$_{-0.67}^{+0.86}$,     11.6$_{-1.5}^{+1.4}$, and    40.83$_{-0.95}^{+0.87}$ \AA,\ respectively (Table~\ref{tab:modelBM1}).
 The broadest component dominates, containing around 68\% of the total H$\beta$ line flux.


\begin{table*}[h]
        \caption{Line fluxes of the most prominent lines visible in the optical spectrum of M1 as fitted in the narrower window with the best coverage of the Fe II templates (model B). }
        \label{tab:modelBM1}
    \centering
\begin{tabular}{llllll}
\hline
 Line      & Wavelength   & Shift    & FWHM      & Flux    & $W_{\lambda}$         \\
&  [\AA]          & [km/s]   &  [km/s]   & [$10^{-17}$ erg s$^{-1}$ cm$^{-2}$]  &  [\AA]     \\
\hline
 Fe II     & 4434-4684               & -290$_{-18}^{+20}$    & 900 $\pm$ 250    & 216 $\pm$ 16                               & 7.56$_{-0.56}^{+0.54}$    \\
 Fe II     & 4434-4684               & 64.6$_{-6.8}^{+3.6}$  & 2250 $\pm$ 250   & 1977$_{-19}^{+20}$                         & 69.31$_{-0.70}^{+0.74}$   \\
 H$\beta$  & 4861.3                  & -2.4$_{-5.4}^{+4.4}$  & 157.0 $\pm$ 8.9      & 20.5 $\pm$ 1.6                             & 0.817$_{-0.061}^{+0.064}$ \\
 H$\beta$  & 4861.3                  & -337 $\pm$ 16         & 722$_{-30}^{+33}$    & 157$_{-17}^{+22}$                          & 6.24$_{-0.67}^{+0.86}$    \\
 H$\beta$  & 4861.3                  & 309$_{-58}^{+75}$     & 1397$_{-68}^{+59}$   & 292$_{-37}^{+35}$                          & 11.6$_{-1.5}^{+1.4}$      \\
 H$\beta$  & 4861.3                  & -338$_{-27}^{+26}$    & 3128$_{-33}^{+37}$   & 1029$_{-25}^{+22}$                         & 40.83$_{-0.95}^{+0.87}$   \\
 H$\gamma$ & 4340.4                  & -343$_{-18}^{+22}$    & 629$_{-41}^{+37}$    & 49.5$_{-7.3}^{+8.1}$                       & 1.57$_{-0.23}^{+0.26}$    \\
 H$\gamma$ & 4340.4                  & 2757$_{-200}^{+120}$  & 1052$_{-130}^{+150}$ & 35.7$_{-8.3}^{+14.0}$                      & 1.16$_{-0.27}^{+0.46}$    \\
 H$\gamma$ & 4340.4                  & 801$_{-47}^{+54}$     & 1481$_{-120}^{+110}$ & 246$_{-41}^{+32}$                          & 7.9$_{-1.3}^{+1.1}$       \\
 H$\gamma$ & 4340.4                  & -425$_{-210}^{+260}$  & 2371$_{-170}^{+180}$ & 155$_{-23}^{+34}$                          & 4.91$_{-0.73}^{+1.10}$    \\

 [O III]   & 5006.843                & -78.7 $\pm$ 3.3       & 199.8 $\pm$ 8.3      & 54.7 $\pm$ 3.7                             & 2.30 $\pm$ 0.16           \\

 [O III]   & 5006.843                & -407 $\pm$ 13         & 503 $\pm$ 12         & 121.9 $\pm$ 4.5                            & 5.12 $\pm$ 0.19           \\
 
\hline
\end{tabular}

\end{table*}

The H$\gamma$ line decomposition is potentially less accurate because of strong Fe II and O III $\lambda$4364 \AA\ contamination which is emphasized by the larger uncertainties on line parameters in comparison to H$\beta$. The FWHM values of its components  in the fit are 629$_{-41}^{+37}$,     1050$_{-130}^{+150}$,     1480$_{-120}^{+110}$, and     2370$_{-170}^{+180}$ km s$^{-1}$
with shifts of -343$_{-18}^{+22}$,     2760$_{-200}^{+120}$,      801$_{-47}^{+54}$,    and -430$_{-210}^{+260}$ km s$^{-1}$,
respectively, while $W_{\lambda}$ values are 1.57$_{-0.23}^{+0.26}$,     1.16$_{-0.27}^{+0.46}$,      7.9$_{-1.3}^{+1.1}$, and     4.91$_{-0.73}^{+1.10}$\AA.
 The fit did not manage to address the profile spike with its width of $\approx$ 200 km s$^{-1}$ as seen in H$\beta$ and [O III] profiles.
The narrowest fitted component of H$\gamma$ does not correspond to the narrowest H$\beta$ Gaussian, and is roughly similar to the second H$\beta$ Gaussian, but lower by 100 km s$^{-1}$, albeit with a very similar blueshift of around -340 km s$^{-1}$. The second H$\gamma$ Gaussian does not correspond to any of the H$\beta$ components. In light of its high redshift, it is more likely to be associated with O III $\lambda$4364, although redshifted as well. The third component with the FWHM value of 1481$_{-120}^{+110}$ km s$^{-1}$ is possibly similar to the H$\beta$ component of corresponding width, but its shift is larger by almost 500 km s$^{-1}$. The broadest component is around 800 km s$^{-1}$ narrower with respect to the broadest H$\beta$ Gaussian and its shift is consistent within the error range.

The [O III] $\lambda$5007 \AA\ line was fitted with FWHMs of 199.8 $\pm$ 8.3 and 503 $\pm$ 12 km s$^{-1}$ and shifts of -78.7 $\pm$ 3.3, and    -407 $\pm$ 13 km s$^{-1}$, where $W_{\lambda}$ is 2.30 $\pm$ 0.16 and    5.12 $\pm$ 0.19 \AA.
The narrowest component of this forbidden line is wider by 30 km s$^{-1}$ and is more blueshifted by 75 km s$^{-1}$. The broadest [O III] component on the other hand is narrower by 219 km s$^{-1}$ and is also blueshifted by 70 km s$^{-1}$. The two-component fit does not fully address the flux excess in the long blue tail of the line, which could be achieved with a third Gaussian. This would require decreasing the broader component FWHM by 100 km s$^{-1}$ and introducing a broader component with a width of around 600 km s$^{-1}$ with a similar blueshift.

The  FWHMs of the H$\beta$ components of the  M2 spectrum are  454$_{-32}^{+33}$,  718 $\pm$ 17,   2155.5$_{-4.9}^{+2.3}$, and     4770$_{-140}^{+110} $km s$^{-1}$ with fitted shifts of 825$_{-24}^{+23}$,   -251.3$_{-7.8}^{+9.6}$,       26$_{-15}^{+16}$, and     -654$_{-31}^{+49}$ km s$^{-1}$, where $W_{\lambda}$ is 1.18 $\pm$ 0.15,     8.88$_{-0.34}^{+0.38}$,     34.7$_{-1.3}^{+1.0}$, and    29.26$_{-0.63}^{+0.65}$\AA,\ respectively (Table~\ref{tab:modelBM2}).
The narrowest peak was not fitted in this case, but rather a small component  of 454 km s$^{-1}$ in width appeared to account for the more prominent red-wing asymmetry with the shift reaching a high value of 825 km s$^{-1}$. 
The component with an FWHM of 718 km s$^{-1}$ is very similar to the corresponding M1 H$\beta$ component with roughly consistent blueshifts, that is, different by 50 km s$^{-1}$. The third component, with a width of 2155.5 km s$^{-1}$, does not have a direct analog in the M1 fit, but is rather found in between the intermediate and broad components of M1 and its shift is also intermediate with respect to those components in M1. 
The decomposition is different from the M1 case as well, given the appearance of the very broad component found significantly above 4000 km s$^{-1}$ with a high blueshift of above 650 km s$^{-1}$. It is particularly worth noting that the residuals after the H$\beta$ fit subtraction from the M2 spectrum are larger on the line blue part --- considerably larger than in the M1 case, as we can see in Fig.~\ref{fig:fitBM2}  (gray line). This is also the case with the SALT spectrum, which is strongly affected by the atmospheric absorption. We therefore think that M2 is also affected by the O$_2$ absorption as we can see in Fig.~\ref{fig:absorption}. As a result, the H$\beta$ profile is distorted and it is difficult to see whether it has changed 
intrinsically.

The [O III] components in the M2 have FWHMs of 195.1 $\pm$ 7.5 and     475$_{-13}^{+12}$ km s$^{-1}$, where shifts are -79.5 $\pm$ 3.6  and   -412 $\pm$ 13 km s$^{-1}$, respectively. This is fully consistent with the M1 model B fit.

\begin{table*}[h]
    \caption{Line fluxes of the most prominent lines visible in the optical spectrum of M2 as fitted in the narrower window with the best coverage of the Fe II templates (model B). }
\label{tab:modelBM2}
    \centering
\begin{tabular}{llllll}
\hline
 Line      & Wavelength   & Shift    & FWHM      & Flux    & $W_{\lambda}$         \\
&  [\AA]          & [km/s]   &  [km/s]   & [$10^{-17}$ erg s$^{-1}$ cm$^{-2}$]  &  [\AA]     \\
\hline
 Fe II     & 4434-4684               & -344 $\pm$ 13          & 900 $\pm$ 250      & 309 $\pm$ 14                               & 11.42$_{-0.52}^{+0.50}$ \\
 Fe II     & 4434-4684               & 197.3$_{-7.9}^{+3.8}$  & 2250 $\pm$ 250     & 1841$_{-19}^{+20}$                         & 68.07$_{-0.80}^{+0.83}$ \\
 H$\beta$  & 4861.3                  & 825$_{-24}^{+23}$      & 454$_{-32}^{+33}$      & 27.9$_{-3.4}^{+3.6}$                       & 1.18 $\pm$ 0.15         \\
 H$\beta$  & 4861.3                  & -251.3$_{-7.8}^{+9.6}$ & 718 $\pm$ 17           & 211.3$_{-8.1}^{+9.0}$                      & 8.88$_{-0.34}^{+0.38}$  \\
 H$\beta$  & 4861.3                  & 26$_{-15}^{+16}$       & 2155.5$_{-4.9}^{+2.3}$ & 824$_{-30}^{+23}$                          & 34.7$_{-1.3}^{+1.0}$    \\
 H$\beta$  & 4861.3                  & -654$_{-31}^{+49}$     & 4766$_{-140}^{+110}$   & 698$_{-15}^{+16}$                          & 29.26$_{-0.63}^{+0.65}$ \\
 H$\gamma$ & 4340.4                  & -355$_{-13}^{+20}$     & 578 $\pm$ 49           & 37.9$_{-8.3}^{+9.7}$                       & 1.26$_{-0.28}^{+0.33}$  \\
 H$\gamma$ & 4340.4                  & 593$_{-130}^{+120}$    & 1318$_{-260}^{+350}$   & 95$_{-44}^{+120}$                          & 3.2$_{-1.5}^{+3.9}$     \\
 H$\gamma$ & 4340.4                  & 561$_{-49}^{+89}$      & 2215$_{-120}^{+150}$   & 292$_{-110}^{+40}$                         & 9.8$_{-3.7}^{+1.4}$     \\
 H$\gamma$ & 4340.4                  & -648$_{-53}^{+110}$    & 3934$_{-260}^{+160}$   & 101$_{-13}^{+14}$                          & 3.35$_{-0.43}^{+0.46}$  \\

 [O III]   & 5006.843                & -79.5 $\pm$ 3.6        & 195.1 $\pm$ 7.5        & 54.8 $\pm$ 3.4                             & 2.45$_{-0.15}^{+0.16}$  \\

 [O III]   & 5006.843                & -412 $\pm$ 13          & 475$_{-13}^{+12}$      & 107.7 $\pm$ 4.2                            & 4.80 $\pm$ 0.19         \\
 
\hline
\end{tabular}

\end{table*}

A closer look at the differences in the spectra level between M1 and M2 as seen on Fig.~\ref{fig:absorption} leads us to the conclusion that M2 suffers from a similar effect to the S spectrum. 
Differences are visible mostly in the 3500--4200\AA\ and 4900--5500 \AA\ windows (~5500--6100 \AA\ and ~7000--7200 \AA\ in observed frame) with up to 10\% in flux level, while differences in the other parts are an order of magnitude smaller.
In particular, atmospheric absorption, although not as severe as in the SALT data, is stronger in the second VLT spectrum. The historical weather log for Paranal on the ESO website\footnote{https://www.eso.org/asm/ui/publicLog}  shows the presence of precipitable water vapor, which may explain stronger and very broad absorption in the M2 spectrum. Broad absorption due to intrinsic obscuration in the source is rather excluded as it does not appear directly on the blue side of the prominent emission lines beside the known atmospheric O$_{2}$ feature. The O$_{2}$ may be responsible for the less smooth profile and larger residuals on the blue side of the H$\beta$ line in the M2 spectrum. If indeed flux in the wavelength window around 4630--4640 \AA\ rest frame was absorbed in M2, this may lead to the apparent increase in broadening of the broadest component.

The most severe atmospheric effects manifest themselves in the SALT spectrum. We therefore consider the fit to this data set to be the least accurate. Additionally, we do not have photometry from the same observing period, and so instead use  photometry from SAAO from  two years earlier. As a result, the flux calibration is uncertain, which prevents us from making any firm conclusions or referring to the flux values here. Rather, we try to rely on the $W_{\lambda}$. 
Fitting Fe II was difficult and less accurate because of our exclusion of the spectral windows with the most drastic absorption. We obtained a fit with broadenings of 1750 $\pm$ 250 and 10000 $\pm$ 250 km s$^{-1}$, which is vastly different from the M1 and M2 fits. The fitted model is presented in Fig.~\ref{fig:fitBS}. This can impact the emission lines, although H$\beta$ should be rather moderately impacted because of the mostly flat flux profile under the line in the template of \cite{boroson1992},
 with the exception of the red tail of the broad component.  Generally, the shape of Fe II  is well addressed by the fit outside of atmospheric absorption windows with the exception of blends around 4500 \AA, where the fit has a lower amplitude than the spectrum. This happened on the red end of the blue CCD chip, and therefore the flux is more uncertain.
The FWHMs of the H$\beta$ components  are 431$_{-42}^{+33}$,      849 $\pm$ 37,    2035$_{-15}^{+16}$ , and    4535$_{-110}^{+120}$ km s$^{-1}$ with shift values of -83 $\pm$ 11,     -496$_{-47}^{+49}$,       93 $\pm$ 13, and  -694.7$_{-1.7}^{+3.7}$ km s$^{-1}$. The widths are very similar to the M2 fit, but deviations in the shifts are significant, especially in the two narrower components.

\begin{table*}[h]
    \caption{Line fluxes of the most prominent lines visible in the optical spectrum of SALT as fitted in the narrower window with the best coverage of the Fe II templates (model B). }
\label{tab:modelBS}     
    \centering
\begin{tabular}{llllll}
\hline
 Line      & Wavelength   & Shift    & FWHM      & Flux    & $W_{\lambda}$         \\
&  [\AA]          & [km/s]   &  [km/s]   & [$10^{-17}$ erg s$^{-1}$ cm$^{-2}$]  &  [\AA]     \\
\hline
 Fe II     & 4434-4684               & 108.3$_{-4.6}^{+2.2}$  & 1750 $\pm$ 250    & 1966 $\pm$ 22                              & 61.95$_{-0.65}^{+0.68}$      \\
 Fe II     & 4434-4684               & 470.4$_{-1.9}^{+26.0}$ & 10000 $\pm$ 250   & 1117$_{-29}^{+26}$                         & 35.19$_{-0.94}^{+0.84}$      \\
 H$\beta$  & 4861.3                  & -83 $\pm$ 11           & 431$_{-42}^{+33}$     & 59 $\pm$ 13                                & 1.89 $\pm$ 0.41              \\
 H$\beta$  & 4861.3                  & -496$_{-47}^{+49}$     & 849 $\pm$ 37          & 159$_{-17}^{+19}$                          & 5.09$_{-0.52}^{+0.60}$       \\
 H$\beta$  & 4861.3                  & 93 $\pm$ 13            & 2035$_{-15}^{+16}$    & 1076 $\pm$ 23                              & 34.36 $\pm$ 0.71             \\
 H$\beta$  & 4861.3                  & -694.7$_{-1.7}^{+3.7}$ & 4535$_{-110}^{+120}$  & 463$_{-21}^{+19}$                          & 14.76$_{-0.65}^{+0.60}$      \\
 H$\gamma$ & 4340.4                  & 326.8$_{-9.8}^{+12.0}$ & 98$_{-22}^{+27}$      & 1.57$_{-0.32}^{+0.37}$                     & 0.0489$_{-0.0098}^{+0.0120}$ \\
 H$\gamma$ & 4340.4                  & 2189$_{-150}^{+1200}$  & 648$_{-83}^{+68}$     & 18.9$_{-5.4}^{+5.2}$                       & 0.59 $\pm$ 0.17              \\
 H$\gamma$ & 4340.4                  & 2862$_{-650}^{+210}$   & 832$_{-74}^{+110}$    & 31.9$_{-5.4}^{+7.5}$                       & 1.00$_{-0.17}^{+0.24}$       \\
 H$\gamma$ & 4340.4                  & 703$_{-14}^{+13}$      & 1710$_{-22}^{+21}$    & 490.3$_{-6.6}^{+6.0}$                      & 15.30$_{-0.22}^{+0.20}$      \\

 [O III]   & 5006.843                & -81.6$_{-5.0}^{+3.3}$  & 227.1$_{-5.2}^{+6.0}$ & 54.8$_{-3.3}^{+3.8}$                       & 1.76$_{-0.11}^{+0.12}$       \\

 [O III]   & 5006.843                & -388$_{-11}^{+10}$     & 402.3$_{-8.4}^{+7.9}$ & 93.5$_{-3.7}^{+3.5}$                       & 3.00 $\pm$ 0.12              \\
 
\hline
\end{tabular}
\tablefoot{Flux of the narrowest [O III] component was used to normalize the SALT spectrum to the corresponding flux of the MUSE fit. }
\end{table*}

\subsubsection{Model W}

The whole fitting procedure was repeated in a broader spectral window between 3544 and 6020~\AA~ to account for all emission lines. This fit is considered less accurate because the iron templates do not fully cover such a broad spectral range. Fits for  H$\beta$ $\lambda$4861.3~\AA, H$\gamma$ $\lambda$4340.4~\AA, and H$\delta$ $\lambda$4102.89~\AA,~ and forbidden lines [O III] $\lambda$5006.84~\AA, [O III] $\lambda$4959~\AA, [O II] $\lambda$3727.3~\AA, [Ne III] $\lambda$3869~\AA, and [Ne V] $\lambda$3425~\AA~ are presented in Table~\ref{tab:modelWM1} for M1 and Table~\ref{tab:modelWM2} for M2. H$\delta$, [O II], [Ne III], and [Ne V] fits are presented in Fig.~\ref{fig:fitlines}. If not stated otherwise in the text, we use the most accurate, narrower fit for H$\beta$, H$\gamma,$ and [O III] lines for further investigations, while we use broader spectral range fits for H$\delta$, [O II], [Ne III], and [Ne V].

\begin{figure*}
        \centering
        \includegraphics[width=0.99\linewidth]{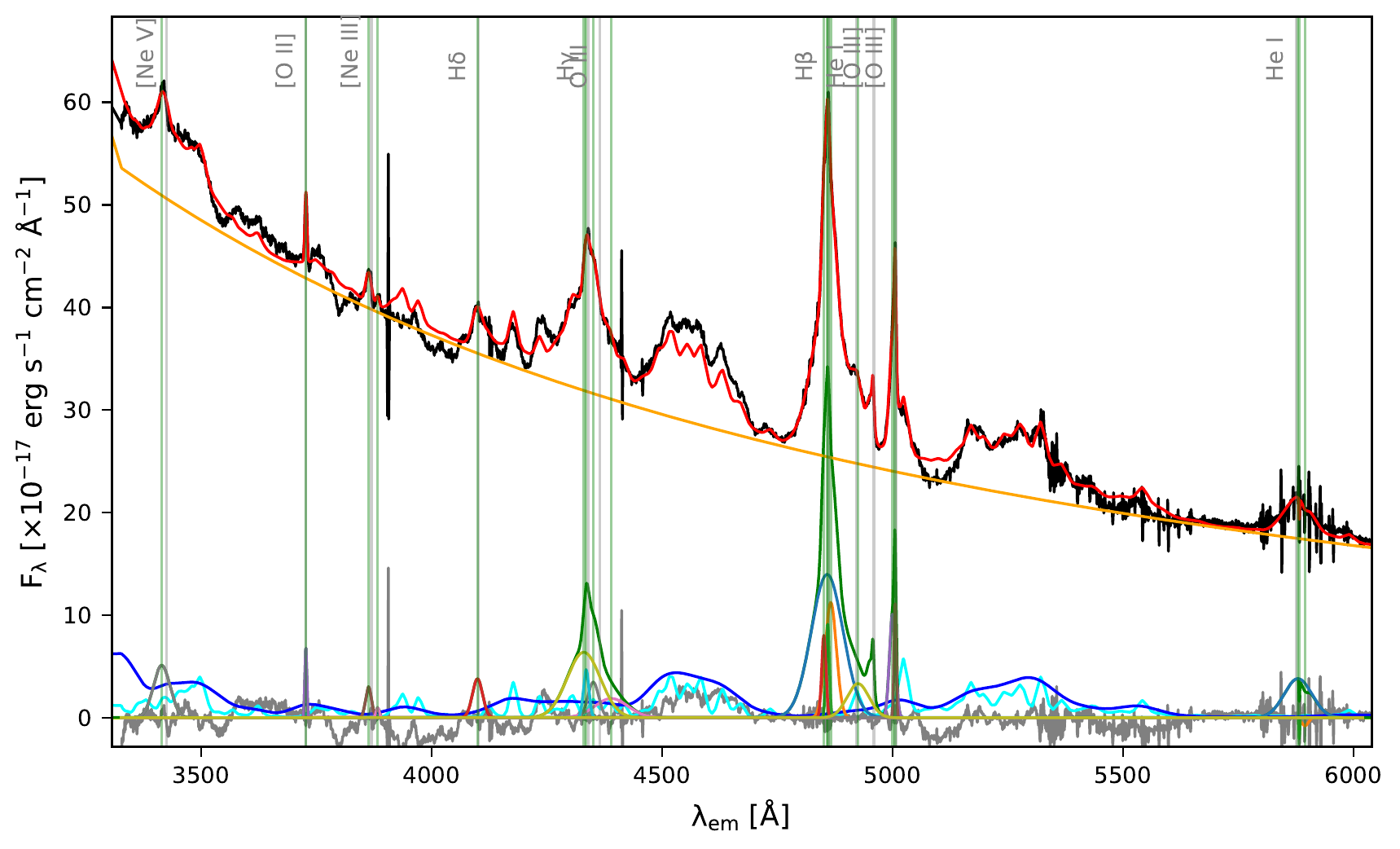}
        \caption{Fit to the M2 spectrum (model W). Color coding is the same as that in Fig.~\ref{fig:fitBM1}.} 
        \label{fig:fitWM2}
\end{figure*}

\begin{figure*}
        \centering
        \includegraphics[width=0.49\linewidth]{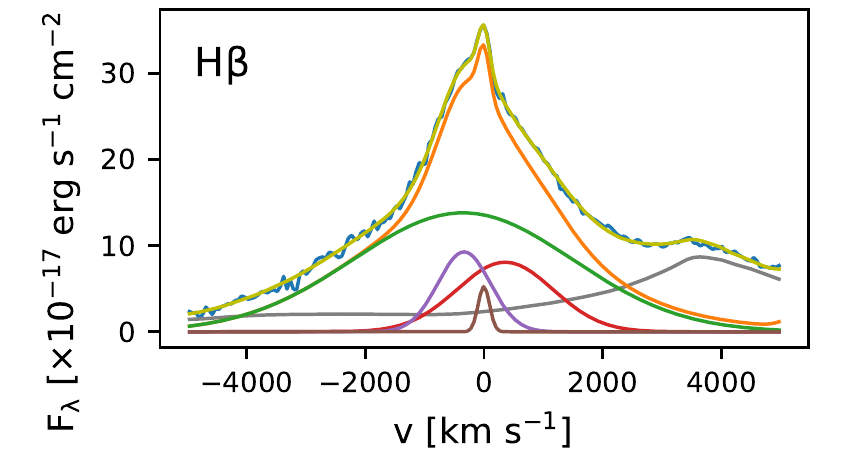}
        \includegraphics[width=0.49\linewidth]{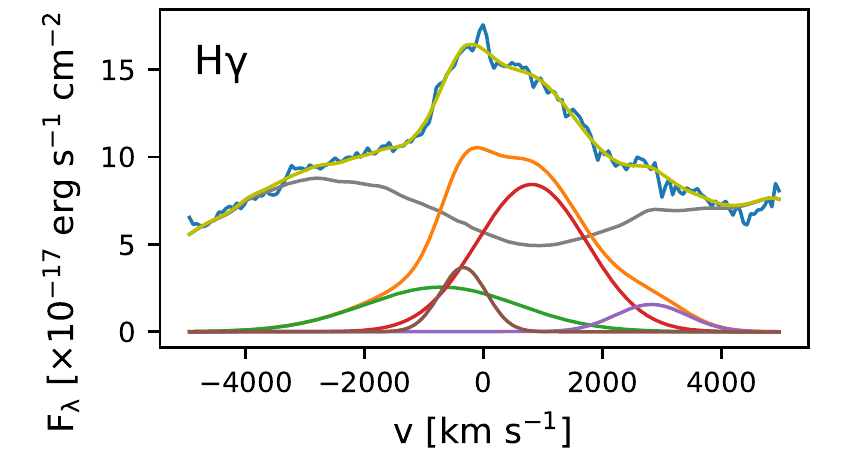}
        \includegraphics[width=0.49\linewidth]{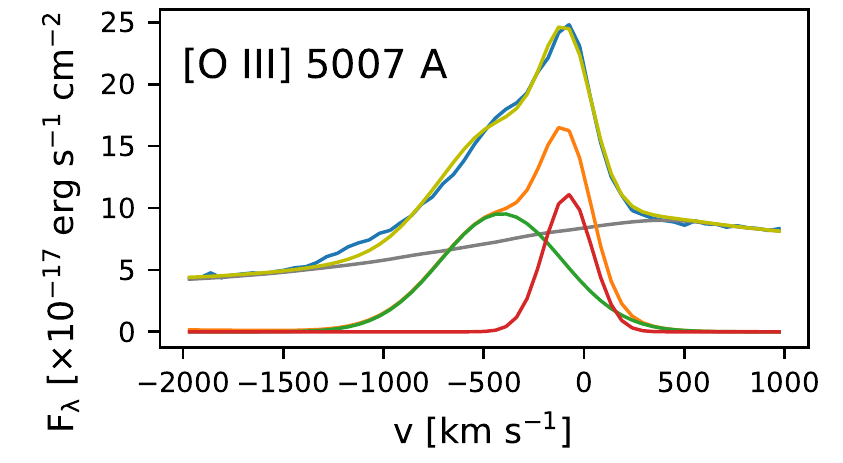}
        \includegraphics[width=0.49\linewidth]{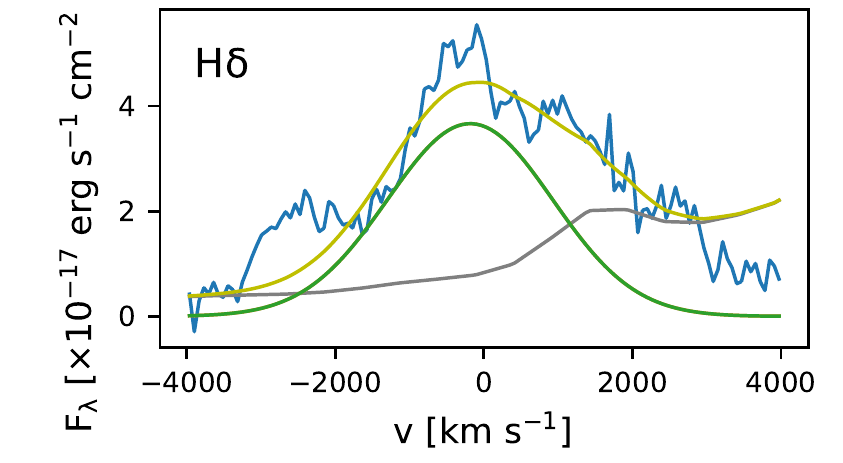}
        \includegraphics[width=0.49\linewidth]{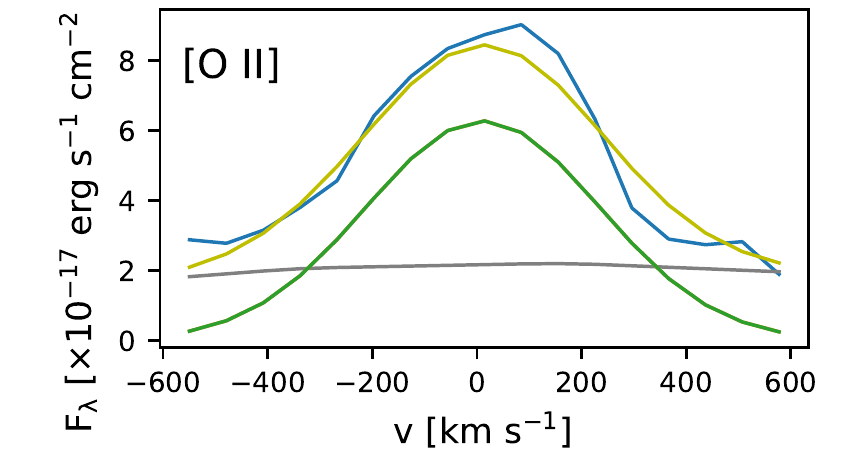}
        \includegraphics[width=0.49\linewidth]{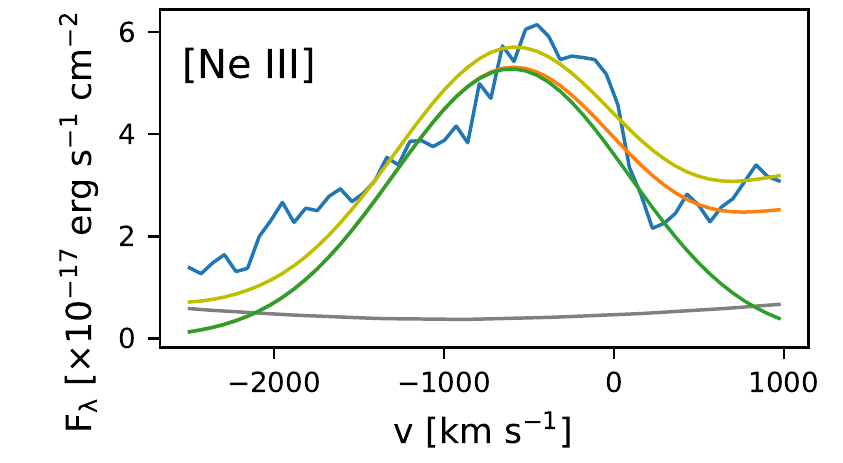}
        \includegraphics[width=0.49\linewidth]{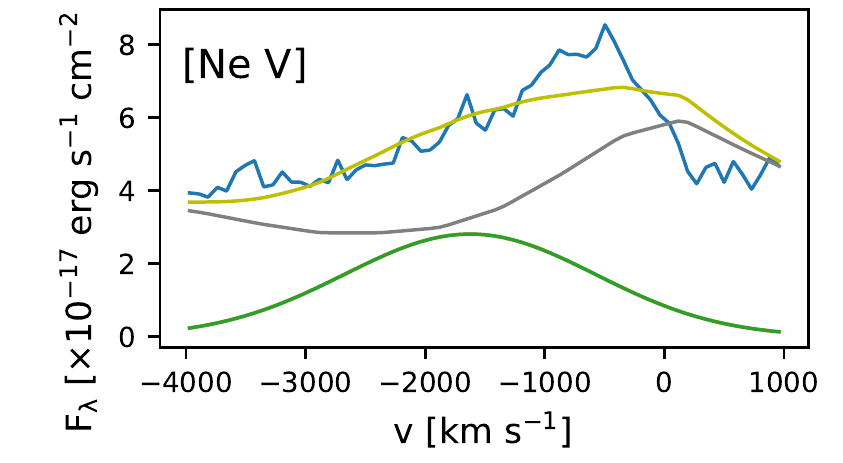}
        
        \caption{ Fitted emission lines (from top left): H$\beta$ $\lambda$4861.3~\AA, H$\gamma$ $\lambda$4340.4~\AA, [O III] $\lambda$5006.84~\AA\ from model B and H$\delta$ 
                $\lambda$4101.73~\AA, [O II] $\lambda$3727.3~\AA, [Ne III] $\lambda$3869~\AA\  from model W. The M1 spectrum is plotted in blue, while the model is presented as a yellow line for the sum of all components (Fe II and Gaussians) in the model.  The sum of Gaussian components is shown in orange, and the separate Gaussian
components  are shown in green, red, violet, and brown. The Fe II template is shown in gray. H$\gamma$ $\lambda$4340.4~\AA\  is contaminated by the  [O III] line, which is potentially the source of the most redshifted component (violet line in the upper right panel). }
        \label{fig:fitlines}
\end{figure*}

The continuum fit was achieved with the power index of -2.0493$_{0.0020}^{0.0024}$ and -1.96637$_{0.0023}^{0.0034}$, where the amplitudes, $F_{5100\text{\AA}}$, are 1.700$_{0.013}^{0.018}$ and 1.621$_{0.025}^{0.015}$ $\times 10^{-16}$ erg s$^{-1}$ cm$^{-2}$ \AA$^{-1}$ for M1 and M2 respectively.
The theoretical template by \cite{bruhweiler2008} in this broadest spectral range fit underestimates the Fe II blends in 3550--3750 \AA\ and 4434--4684 \AA\ wavelength ranges. Also, in the case of M2, this latter template overestimates the region of 3800--4050 \AA.
Thesew ranges are therefore not used further except for [O II]-based relations.
We give fitted values of the W model for the emission lines of M1 and M2 in Tables~\ref{tab:modelWM1} and \ref{tab:modelWM2}, and the example of M2 fit is presented in Fig.~\ref{fig:fitWM2}.

In the M1 case, the fitted narrow Fe II template uses simulation parameters, hydrogen number density, log ($n_H$ [cm$^{-3}$]) = 11.5, 
hydrogen ionizing photon flux,
log ($\Phi(\mathrm{H})$ [cm$^{-2}\, \mathrm{s}^{-1}$]) = 20.5, and microturbulence, $\xi$ = 20 km s$^{-1}$
with FWHM = 1500 $\pm$ 200 km s$^{-1}$.
While the broad component log ($n_H$ [cm$^{-3}$]) = 11, log ($\Phi(\mathrm{H})$ [cm$^{-2}\, \mathrm{s}^{-1}$]) = 20 and $\xi$ = 20 km s$^{-1}$ is broadened to
FWHM = 2500 $\pm$ 200 km s$^{-1}$.
In the case of M2, the parameters were log ($n_H$ [cm$^{-3}$]) = 11 and log ($\Phi(\mathrm{H})$ [cm$^{-2}\, \mathrm{s}^{-1}$]) = 21, $\xi$ = 20 km s$^{-1}$ with broadening of 1500 $\pm$ 200 km s$^{-1}$. For the broader component, the template with log ($n_H$ [cm$^{-3}$]) = 11, log ($\Phi(\mathrm{H})$ [cm$^{-2}\, \mathrm{s}^{-1}$]) = 20.5, and $\xi$ = 50 km s$^{-1}$ was used with broadening of
2900 $\pm$ 200 km s$^{-1}$.
As the iron blend fit was not accurate, we do not interpret those parameters as physically meaningful. However, the fit provides a rough idea of the Fe II contamination of the forbidden lines outside the model B fitting window, namely [O II], [Ne III], [Ne V], and the Balmer H$\delta$ line, and enables qualitative inspection of the profiles.

In comparison to model B, model W results in a higher level of continuum but with a similar power index. This resulted in a lower amplitude of the Fe II emission, especially under the H$\beta$ line profile. As a consequence of a weaker iron fit, and the fact that the theoretical template does not  fully predict the feature around 4916 \AA\  between H$\beta$ and the [O III] complex, an additional component of He I 4921 \AA\  was fitted. Overall, H$\beta$ components became weaker and narrower in comparison with model B. As the model W fits are much more accurate due to the strong limitations of the Fe II fit, as is clearly visible in Fig.~\ref{fig:fitWM2}, we do not refer to this model any further, except to discuss [O II], [Ne III], [Ne V], and Balmer H$\delta$.

\begin{table*}[h]
    \caption{Line fluxes of the most prominent lines visible in the optical spectrum of M1 as fitted in the broad window covering most of the spectrum (model W). }
\label{tab:modelWM1}
    \centering
\begin{tabular}{llllll}
\hline
 Line      & Wavelength   & Shift    & FWHM      & Flux    & $W_{\lambda}$         \\
&  [\AA]          & [km/s]   &  [km/s]   & [$10^{-17}$ erg s$^{-1}$ cm$^{-2}$]  &  [\AA]     \\
\hline
 Fe II     & 4434-4684               & -499.10$_{-0.47}^{+0.54}$ & 1500 $\pm$ 200     & 415.4$_{-7.7}^{+5.7}$                      & 13.62$_{-0.27}^{+0.20}$      \\
 Fe II     & 4434-4684               & -998.63$_{-0.36}^{+0.93}$ & 2500 $\pm$ 200     & 925.74$_{-0.93}^{+0.29}$                   & 30.226$_{-0.014}^{+0.054}$   \\
 H$\beta$  & 4861.3                  & -5.1$_{-3.1}^{+9.3}$      & 155$_{-12}^{+5}$       & 20.2$_{-1.3}^{+1.2}$                       & 0.755$_{-0.047}^{+0.041}$    \\
 H$\beta$  & 4861.3                  & -321$_{-15}^{+4}$         & 723.4$_{-3.7}^{+43.0}$ & 191$_{-25}^{+9}$                           & 7.12$_{-0.89}^{+0.35}$       \\
 H$\beta$  & 4861.3                  & 502$_{-96}^{+18}$         & 1274$_{-66}^{+95}$     & 217$_{-26}^{+49}$                          & 8.14$_{-0.96}^{+1.90}$       \\
 H$\beta$  & 4861.3                  & -265$_{-77}^{+34}$        & 3009$_{-27}^{+29}$     & 1042$_{-40}^{+21}$                         & 38.8$_{-1.5}^{+0.8}$         \\
 H$\gamma$ & 4340.4                  & -351$_{-16}^{+7}$         & 661$_{-32}^{+26}$      & 58.7$_{-9.0}^{+14.0}$                      & 1.73$_{-0.27}^{+0.41}$       \\
 H$\gamma$ & 4340.4                  & 652$_{-110}^{+140}$       & 1279$_{-140}^{+210}$   & 106$_{-12}^{+24}$                          & 3.15$_{-0.34}^{+0.70}$       \\
 H$\gamma$ & 4340.4                  & -696$_{-14}^{+4200}$      & 4121$_{-22}^{+9}$      & 446$_{-250}^{+19}$                         & 13.1$_{-7.2}^{+0.6}$         \\
 H$\gamma$ & 4340.4                  & 3393$_{-4200}^{+26}$      & 4131$_{-12}^{+11}$     & 206.9$_{-5.7}^{+270.0}$                    & 6.27$_{-0.18}^{+7.60}$       \\
 H$\delta$ & 4101.73                 & -197$_{-52}^{+14}$        & 1851$_{-22}^{+68}$     & 146.8$_{-7.4}^{+1.5}$                      & 3.86$_{-0.19}^{+0.04}$       \\
 He I      & 5875.6                  & 213.1$_{-7.4}^{+8.6}$     & 2543$_{-23}^{+4}$      & 243.5$_{-7.1}^{+2.1}$                      & 13.41$_{-0.38}^{+0.12}$      \\

 [O III]   & 5006.843                & -85.8$_{-2.6}^{+4.8}$     & 211.2$_{-3.6}^{+8.8}$  & 58.6$_{-2.3}^{+1.7}$                       & 2.325$_{-0.092}^{+0.062}$    \\

 [O III]   & 5006.843                & -410.2$_{-4.3}^{+11.0}$   & 562$_{-16}^{+20}$      & 143.1$_{-1.4}^{+4.6}$                      & 5.666$_{-0.062}^{+0.180}$    \\

 [O II]    & 3727.3                  & 11.4$_{-3.9}^{+2.5}$      & 354.4$_{-9.3}^{+18.0}$ & 43.81$_{-0.38}^{+0.61}$                    & 0.9488$_{-0.0077}^{+0.0140}$ \\

 [Ne III]  & 3869                    & -617$_{-42}^{+21}$        & 1150$_{-13}^{+10}$     & 114.8$_{-5.1}^{+3.1}$                      & 2.67$_{-0.12}^{+0.08}$       \\

 [Ne V]    & 3425                    & -1619$_{-46}^{+58}$       & 1874$_{-140}^{+86}$    & 86.8$_{-2.8}^{+5.1}$                       & 1.564$_{-0.048}^{+0.091}$    \\
 
\hline
\end{tabular}

\end{table*}

\begin{table*}[h]
    \caption{Line fluxes of the most prominent lines visible in the optical spectrum of M2 as fitted in the broad window covering most of the spectrum (model W). }
\label{tab:modelWM2}    
    \centering
\begin{tabular}{llllll}
\hline
 Line      & Wavelength   & Shift    & FWHM      & Flux    & $W_{\lambda}$         \\
           &  [\AA]       & [km/s]   &  [km/s]   & [$10^{-17}$ erg s$^{-1}$ cm$^{-2}$]  &  [\AA]     \\
\hline
 Fe II     & 4434-4684               & -497.86$_{-0.40}^{+0.55}$ & 1500 $\pm$ 200           & 469.9$_{-6.8}^{+3.1}$                      & 16.30$_{-0.24}^{+0.13}$    \\
 Fe II     & 4434-4684               & -971$_{-16}^{+3}$         & 2900$_{-200}^{+200}$ & 814.75$_{-0.65}^{+0.16}$                   & 28.226$_{-0.022}^{+0.040}$ \\
 H$\beta$  & 4861.3                  & -131$_{-16}^{+19}$        & 367$_{-13}^{+7}$             & 83.5$_{-5.1}^{+2.4}$                       & 3.29$_{-0.21}^{+0.09}$     \\
 H$\beta$  & 4861.3                  & -655$_{-18}^{+20}$        & 453$_{-47}^{+9}$             & 91$_{-15}^{+9}$                            & 3.56$_{-0.57}^{+0.36}$     \\
 H$\beta$  & 4861.3                  & 281$_{-14}^{+52}$         & 1246 $\pm$ 36                & 342$_{-28}^{+17}$                          & 13.5$_{-1.2}^{+0.6}$       \\
 H$\beta$  & 4861.3                  & -232$_{-31}^{+33}$        & 3450$_{-54}^{+80}$           & 1194$_{-11}^{+10}$                         & 46.90$_{-0.42}^{+0.46}$    \\
 H$\gamma$ & 4340.4                  & -357.7$_{-10.0}^{+13.0}$  & 681$_{-17}^{+7}$             & 69.6$_{-1.1}^{+5.7}$                       & 2.189$_{-0.035}^{+0.180}$  \\
 H$\gamma$ & 4340.4                  & 709$_{-23}^{+49}$         & 1232$_{-80}^{+94}$           & 91.7$_{-8.3}^{+8.9}$                       & 2.90$_{-0.27}^{+0.29}$     \\
 H$\gamma$ & 4340.4                  & 3415$_{-24}^{+1}$         & 4125$_{-12}^{+4}$            & 177.1$_{-8.6}^{+7.9}$                      & 5.70$_{-0.27}^{+0.25}$     \\
 H$\gamma$ & 4340.4                  & -713.0$_{-3.4}^{+3.8}$    & 4141.7$_{-7.0}^{+0.9}$       & 576.1$_{-2.3}^{+9.6}$                      & 18.068$_{-0.100}^{+0.290}$ \\
 H$\delta$ & 4101.73                 & -134$_{-29}^{+14}$        & 1333$_{-49}^{+19}$           & 103.6$_{-8.3}^{+0.7}$                      & 2.91$_{-0.24}^{+0.02}$     \\
 He I      & 5875.6                  & 197.1$_{-3.7}^{+24.0}$    & 2545.3$_{-1.1}^{+4.8}$       & 287.9$_{-1.6}^{+3.9}$                      & 16.445$_{-0.046}^{+0.230}$ \\

 [O III]   & 5006.843                & -91.3$_{-6.8}^{+3.0}$     & 222$_{-13}^{+14}$            & 62.8$_{-7.4}^{+1.6}$                       & 2.62$_{-0.32}^{+0.06}$     \\

 [O III]   & 5006.843                & -428.5$_{-3.3}^{+28.0}$   & 642$_{-12}^{+4}$             & 164.1$_{-6.4}^{+3.5}$                      & 6.83$_{-0.27}^{+0.14}$     \\

 [O II]    & 3727.3                  & -0.9$_{-3.4}^{+11.0}$     & 286.7$_{-7.6}^{+8.6}$        & 35.90$_{-0.71}^{+1.60}$                    & 0.838$_{-0.018}^{+0.036}$  \\

 [Ne III]  & 3869                    & -457$_{-15}^{+2}$         & 737$_{-37}^{+77}$            & 46.5$_{-3.4}^{+1.3}$                       & 1.163$_{-0.085}^{+0.033}$  \\

 [Ne V]    & 3425                    & -921$_{-130}^{+25}$       & 2616.6$_{-4.3}^{+8.5}$       & 230.0$_{-5.4}^{+0.9}$                      & 4.52$_{-0.12}^{+0.02}$     \\
 
\hline
\end{tabular}

\end{table*}

\subsection{Global parameters}

Here we derive global parameters like SMBH mass, bolometric luminosity, Eddington ratio, and position on the quasar main sequence (QMS) diagram using optical spectra. We rely mostly on M1 as the best-quality spectra with which to derive those parameters and use model B because it is closest to what is used in the literature. While referring to the [O II] line fits we use model W as it incorporates this line.

To characterize the sources within the AGNs population, we use eigen vector 1 (EV1) parameters defined by \cite{boroson1992} and \cite{sulentic2000}. These contain the ratio of iron to H$\beta$ $W_{\lambda}$, the FWHM of H$\beta$, and the [O III] $W_{\lambda}$. The ratio between the emission
strengths of Fe II  and H$\beta$ is defined as
\begin{equation}\label{rfe}
R_{FeII} = \frac{W_{\lambda}(\mathrm{Fe II}, 4434-4684\text{\AA})}{W_{\lambda}(\mathrm{H\beta})}
,\end{equation}
where $W_{\lambda}(\mathrm{Fe II}, 4434-4684\text{\AA})$ is the equivalent width of Fe II integrated over 4434--4684~\AA\ wavelength range and $W_{\lambda}(\mathrm{H\beta})$ is the equivalent width of the H$\beta$ line broad component. Although \cite{boroson1992} and \cite{sulentic2000} mention a broad component of H$\beta$, these authors effectively account for the total profile in type A sources defined as a single Lorentzian.
We obtained a total $W_{\lambda}(\mathrm{H\beta})$ of 59.55$_{-0.33}^{+0.31}$ \AA,\ and for Fe II integrated over 4434--4684~\AA\ $W_{\lambda}$ is 76.88 $\pm$ 0.48~\AA,\  giving $R_{\mathrm{FeII}}$ = 1.2910$_{-0.0070}^{+0.0074}$, which is above the average. 

Our Fe II fit consists of two components, one narrower and one broader. The broad component is fitted with FWHM = 2250 km s$^{-1}$ and a shift of 197.3 km s$^{-1}$ with the template by~\cite{boroson1992} in model B. In general, the preferred broadening is in good agreement with H$\beta$ components if we compare the narrower components, but for broader Fe II, the value is in between intermediate and broad H$\beta$ components in M1, but consistent with the intermediate component width of M2. 
The shifts are different, but as already noted in the literature, both outflow and inflow are present in the emitting region  \citep{czerny2011}   and generally Fe II emission can be shifted with respect to prominent emission lines. 

To compute H$\beta$ BC FWHM in line with the original EV1 parameter definition, we used model B continuum + Fe II and refitted H$\beta$ with a single Lorentzian profile. We obtained a value of 2275$_{-15}^{+13}$ km s$^{-1}$ (2419$_{-14}^{+11}$ km s$^{-1}$ for M2 and 2158.8$_{-5.5}^{+5.9}$ for SALT).

Using the virial black hole (BH) mass formula by \cite{vestergaard2006} we get
1.17$_{-0.43}^{+0.70}$ $\times 10^{8}$ M$_{\odot}$ for the derived L$_{5100\text{\AA}}$ = 1.515 $\pm$ 0.065 $\times 10^{41}$ erg s$^{-1}$ \AA$^{-1}$ of the continuum luminosity
and FWHM(H$\beta$) = 2275$_{-15}^{+13}$ km s$^{-1}$ for the broad component of H$\beta$.
We then estimate bolometric luminosity to be $L_{bol}$ = 7.98 $\pm$ 0.34 $\times 10^{45}$ erg s$^{-1}$ using the~\cite{richards2006} relation with L$_{5100\text{\AA}}$. From the above parameters, we compute the Eddington ratio which is $L_{bol}/L_{Edd}$ = 0.54$_{-0.21}^{+0.32}$.
The uncertainties on the BH mass value are adopted from the work of \citep{mejia2016}  (0.2 dex). The log-normal distribution of masses was then used in computations of the Eddington luminosity and ratio.

Using H$\beta$, [O III], and [O II] fits, we may estimate bolometric luminosity based on relations by \cite{punsly2011}.
The $H\beta$ line-luminosity-based bolometric luminosity value is equal to 5.370$_{-0.098}^{+0.089}$ $\times 10^{45}$  erg s$^{-1}$ and thus $L_{bol}/L_{Edd}$ = 0.193$_{-0.072}^{+0.12}$.
The shape of the [O III] spectral profile agrees with the one shown in Figure 2 in the paper by \cite{shen2014} with log 5100 L$_{5100\text{\AA}}$ = 44.9. 
From the relation with [O III] $\lambda 5007\text{\AA}$, the bolometric luminosity is 7.100 $\pm$ 0.038 $\times 10^{45}$ erg s$^{-1}$ and thus the Eddington ratio is 0.48$_{-0.19}^{+0.29}$. This latter is possibly underestimated because of overestimated Fe II, but is overall consistent with the estimate from the continuum at 5100\AA. While incorporating the relation with [O II] and using model W  we get $L_{bol}$ = $9 \times 10^{45}$ erg s$^{-1}$ and an Eddington ratio of 0.6, which is higher. 
One caveat is the much more uncertain Fe II fit in the [O II] spectral window of model W and a second is the contribution from star formation, which is strong in this source.

Given errors are purely fit errors derived from the solution probability distribution corresponding to a one-sigma confidence level. However, taking into account differences between spectra and fits, the realistic uncertainty on the global parameters would be two orders of magnitude larger.

The Eddington ratio values are moderate. Many authors have suggested that most of the continu-um slope of a type 1 AGNs  deviates from the accretion disk model prediction \citep{shankar2016}, which is due to intrinsic reddening \citep{baron2016}. We performed additional checks to evaluate whether or not our $L_{bol}$ is affected by instrinsic redenning. Although HST spectra do not show signs of distortion caused by intrinsic extinction, we further test this possibility. To look for possible attenuation, we investigated the He I $\lambda$ 5875.6 \AA\ emission line in the search for possible Na I D absorption (as suggested e.g. by \cite{baron2016}). Besides the sky residual features present over the He I spectral profile there is no obvious absorption. If any is present, it is very mild, suggesting very little if any intrinsic extinction affecting the AGN. Additionally, $L_{bol}$ estimations based on the continuum at 5100 \AA\ or [O III] are not expected to be noticeably affected by mild extinction.

\subsection{Host galaxy}

\begin{figure*}
        \centering
        \includegraphics[width=1\linewidth]{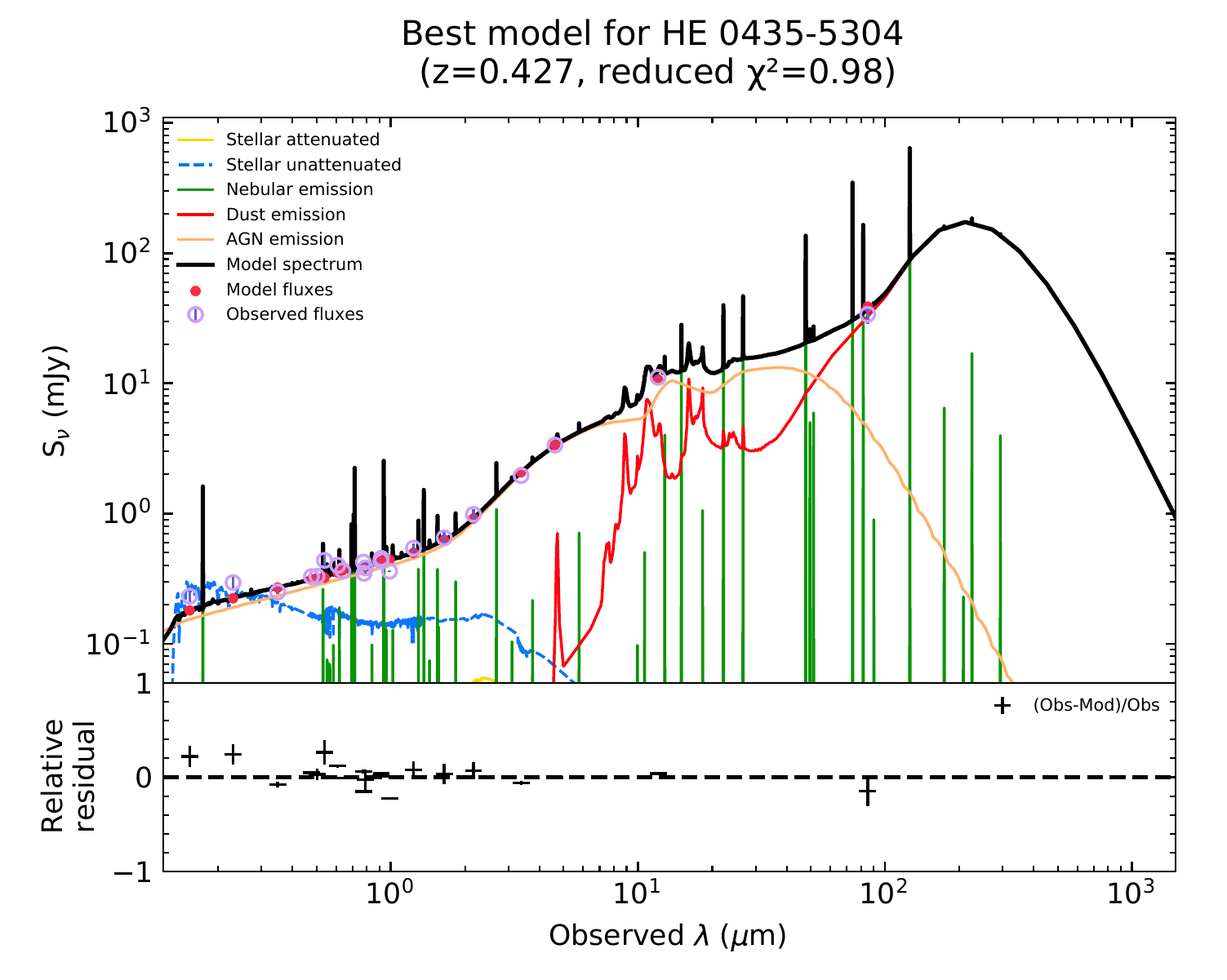}
        \caption{Best fit (in black) of the constructed SEDs of HE~0435-5304 and the relative residuals.  The unattenuated stellar emission is shown with the blue line, while the yellow line shows the attenuated stellar emission. Red line shows the dust emission and the broad line represents the AGN component.  Red dots are the best-fit values of the observations which are shown with  purple open circles. }
        \label{fig:cigale}%
\end{figure*}

The main physical properties of HE~0435-5304 based on the CIGALE SED fitting procedure are listed in~Fig. \ref{tab:cigale}, while Fig.~\ref{fig:cigale} shows the final fit of the SED. We find that HE~0435-5304 is an ULIRG with total dust luminosity ($L_{IR\mbox{ }total}^{*}$) calculated as the sum of the dust luminosity (L$_{dust}$) and the AGN luminosity emitted at IR wavelengths (L$_{dust\mbox{ }AGN}$) equal to $10^{12.14\pm0.11}$ L$_{\odot}$. 

The galaxy is relatively massive ($10^{10.12\pm0.62}$ M$_{\odot}$) and is actively star forming at a rate of SFR=48.90~$\pm$~12.78 M$_{\odot}$/yr, with a specific SFR (defined as SFR over M$_{star}$) of log SSFR=-8.44 Myr$^{-1}$. Based on Eq.~9 from \cite{Schreiber:2015} we calculate the SFR$_{MS}$, the average SFR of the main sequence (MS) galaxies at the same redshift and the same stellar mass as the  HE~0435-5304. We find that the SFR of HE~0435-5304 is more than seven times higher than the average SFR of MS galaxies  ($SFR_{z=0.427, log M_{\star}=10.12}=6.533 M_{\odot}yr^{-1}$), which implies that HE~0435-5304 is significantly more active in star formation than the majority of star forming galaxies at the same redshift, albeit not as active as for example starburst galaxies.

The AGN contribution to the total dust luminosity for HE~0435-5304  was estimated as a 56~$\pm$~6\%. The viewing angle of the AGN is found to be 14.26~$\pm$~10.48 $deg$ which is consistent with HE~0435-5304 being a type 1 AGN. In spite of the fact that LIRGs and ULIRGs are known to be large dust reservoirs, the AGN in this object does not appear to be significantly attenuated and dominates not only the optical but also the IR part of the spectrum. 

\begin{table}[h!]
\caption{Main physical properties of HE~0435-5304 based on CIGALE SED fitting.} 
\label{tab:cigale}
        \centering
\begin{tabular}{ll}
\hline
 Parameter      & Estimated value   \\
\hline
 log$L_{dust}$ [$L_{\odot}$] & 11.79 $\pm$ 0.23   \\
 log$L_{IR\mbox{ }total}^{*}$ [$L_{\odot}$] & 12.14$\pm$ 0.11\\
  log$M_{star}$ [$M_{\odot}$] & 10.12  $\pm$ 0.62     \\
  SFR (100 Myrs) [$M_{\odot} yr^{-1}$] & 48.90 $\pm$ 12.78 \\
  SFR (10 Myrs) [$M_{\odot} yr^{-1}$] & 60.56 $\pm$ 15.76 \\
  AGN frac & 0.56 $\pm$ 0.06 \\
  AGN viewing angle [deg]& 14.26 $\pm$ 10.48 \\
  log(AGN$_{bol}$). lum &  12.57 $\pm$ 0.05 \\
  log(AGN$_{dust}$). lum &  11.89 $\pm$ 0.06 \\ 
  $H_{\beta}$ & 8.24$\times$10$^{-20}$  $\pm$ 1.87$\times$10$^{-19}$ \\
  $O_{II}$  & 4.33$\times$10$^{-20}$  $\pm$ 1.19$\times$10$^{-19}$ \\
  $O_{III}$495.9  & 1.08$\times$10$^{-20}$  $\pm$ 2.41$\times$10$^{-19}$ \\
  $O_{III}$500.7  & 3.29$\times$10$^{-19}$  $\pm$ 7.33$\times$10$^{-19}$ \\
\hline
\end{tabular}

\end{table}

\begin{figure}
        \centering
        \includegraphics[width=0.99\linewidth]{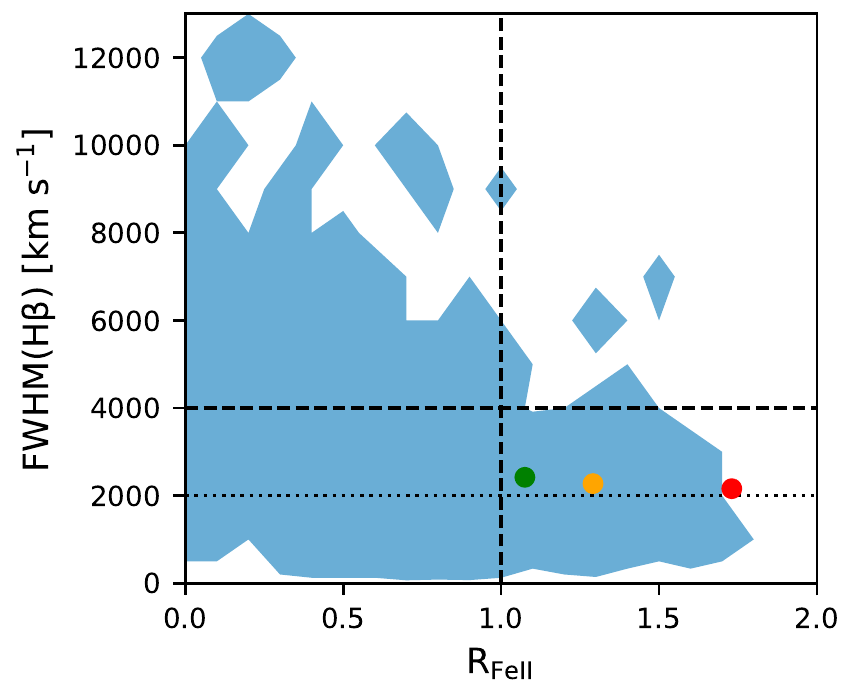}
        \caption{QMS diagram with the position of HE~0435-5304  according to the fitted parameters  overplotted. The blue shaded area shows the AGNs population \citep[based on the][catalog]{shen2011}, while orange, green, and red points show the position of  HE~0435-5304 as  derived from M1, M2, and SALT spectra, respectively. The points were computed assuming a single Lorentzian component fit in H$\beta$ and continuum + Fe II from model B. The horizontal dashed line splits the population into groups of type A (bottom) and type B (top), while the vertical dashed line divides AGNs into normal (left) and extreme (right). The dotted horizontal line divides population A into NLSy1 (FWHM $<$ 2000 km s$^{-1}$) and sources with broader lines.
        }
        \label{fig:qms}%
\end{figure}

\begin{figure}
        \centering
        \includegraphics[width=0.99\linewidth]{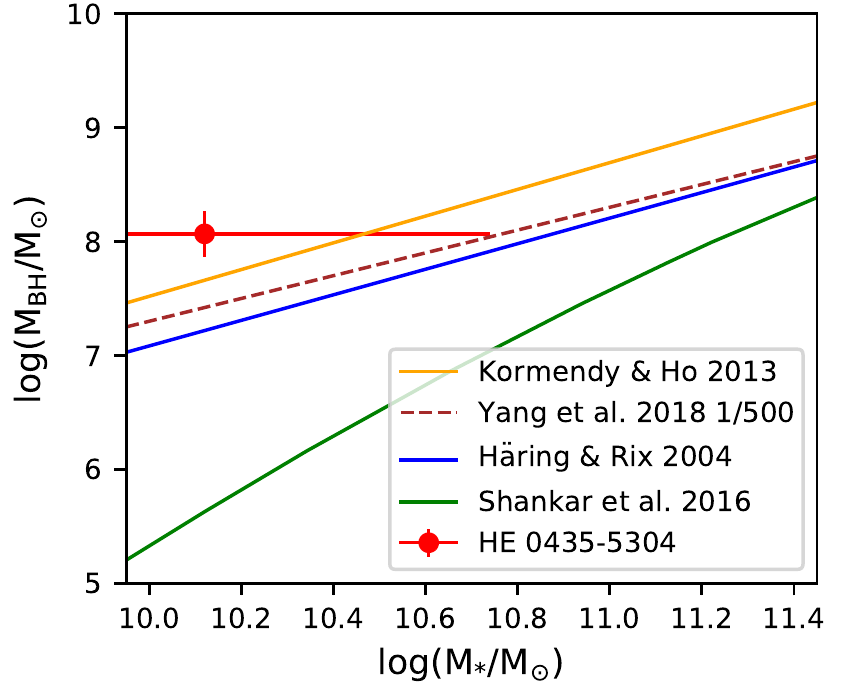}
        \caption{Relation between stellar mass and mass of the central black hole as was presented by \cite{yang2018} for galaxies with broad lines. The orange line presents the relation by \cite{kormendy2013}, the brown dashed line is the relation $M_{*}/M_{BH}$=1/500 by \cite{yang2018}, the blue line follows results by \cite{haring2004}, and the green line is based on work by \cite{shankar2016a}.
        }
        \label{fig:mstarmbh}%
\end{figure}

\section{Discussion}

Generally, HE 0435-5304 is a quasar with prominent lines of moderate width. Its H$\beta$ FWHM is of the order of 3200 km s$^{-1}$ which makes it a type A source on the QMS.
Its very prominent Fe II emission makes it a member of the extreme population A (xA).  Population A is known to have narrower Lorentzian-like lines with systematically higher accretion rates, while population B is characterized by broader overall Gaussian-like H$\beta$ line profiles and lower accretion rates \citep{sulentic2000,marziani2018}.
Although the overall H$\beta$ profile is only roughly Lorentzian-like, excellent VLT spectra reveal asymmetries: blue excess  near the core, an intermediate red tail, and a far blue tail. These asymmetries can be fully addressed by four Gaussian components with shifts in the range -300--300 km s$^{-1}$.
When we consider the [O III] $\lambda$5007~\AA\ line strength and the $R_{FeII}$ value, our source is located in the type A region. The general appearance of the line profiles with their asymmetry and luminosity is also consistent with a Population A source \citep{shen2014}. When we place our source on the EV1 plane parameters, FWHM(H$\beta$), $R_{FeII}$ , and log $W_{\lambda}$([O III]), all values are consistent with the population presented in Figure 1 in the article by \cite{shen2014}. 
As noted by \cite{shen2014}, $R_{FeII}$ correlates with the hot dust luminosity at IR. HE 0435-5304 is IR-bright as it belongs to the ULIRG group, and therefore its relatively strong iron emission is not surprising. It is worth noting that with an SMBH mass of 1.17$_{-0.43}^{+0.70}$ $\times 10^{8}$ M$_{\odot}$, we get a relatively high Eddington ratio of 0.54$_{-0.21}^{+0.32}$ which is consistent with the population A sources as derived from the M1 B fit. However, considering its $R_{FeII}$ , which is well above 1, HE 0435-5304 resides in the extreme type A region \citep{marziani2018} on the QMS diagram, as presented on Fig.~\ref{fig:qms} \citep[similar to][]{zamfir2010}. The QMS is plotted in Fig. 10 based on the SDSS DR7 spectral catalog by \cite{shen2011}.
The CIGALE fit of the SED gives us an estimated AGN luminosity of between $4.6 \times 10^{45}$ and $7.1 \times 10^{45}$ erg s$^{-1}$ with its strong IR emission which is consistent with values obtained from the spectra. The result based on the full IR-to-UV range is roughly consistent 
with the value inferred from spectral fit alone ($L_{bol}$ = 7.98 $\pm$ 0.34 $\times 10^{45}$ erg s$^{-1}$ from the 5100 \AA\ continuum luminosity or 5.370$_{-0.098}^{+0.089}$ $\times 10^{45}$ erg s$^{-1}$ from the H$\beta$ line luminosity).

The M2 spectrum  was taken on a night with over 80\% higher relative humidity in comparison to M1 and precipitable water vapor present in the air. The effect of this manifests itself mostly in similar windows to those of the SALT spectrum (see Fig.~\ref{fig:absorption}), namely distortion blueward of H$\gamma$ and redward of [O III] 5007 \AA\ lines.
The consequence is not surprising: poorer weather conditions give higher true uncertainties. 
Fit parameters
are influenced as a result, although not as severely as in the SALT spectrum. The resulting parameters have larger uncertainties in continuum  and Fe II components (probably
over an order of magnitude larger) than those given by the fit statistics itself. This may influence the emission line fits, although in the case of the M2 spectrum, we do not expect this to impact the conclusions presented here. However, in the case of the SALT spectrum, 
in spite of observational problems, the intrinsic physical effect may still be significant. Over time, the broadest component increases in width together with the decline  in source luminosity, as we can see in the photometric light curves. Also, we know that lower luminosity means lower radiation pressure on the inner face of BLR, which results in it moving inward, closer to the SMBH. Therefore, the  intrinsic change to the H$\beta$ profile is possible, considering the difference in the observation dates between M1 and M2 of 3 months, at least in the broad and possibly in the intermediate components, but probably not in the narrowest one.

Considering population A, we expect a lower variability amplitude in comparison to objects with broader lines and lower accretion rates. This is the case for HE 0435-5304.  Figure~\ref{fig:lightcurve} shows that the variability of the amplitude on a timescale of tens of days is low. Instead, in the Catalina light curve, we see a rather steady long-term decline over the years which seems to be similar to for example the case of HE 0413-4031 \citep{zajacek2020}.
The level of the UV flux visible in the HST spectra from 2003 and 2010 is roughly similar (Fig.\ref{fig:spectra}, left panel). Although photometry in the FUV filter derived from HST data shows a trend consistent with the overall decline over time with GALEX FUV point in between.

HE 0435-5304 demands a two-component fit of the Fe II blended emission with clearly visible narrow peaks. This is consistent with the theoretical radial emissivities for Fe II, where we expect prominent emission from both broad and intermediate line regions \citep{adhikari2018}. With broadening of the iron emission at 900 and 2250 km s$^{-1}$ the fit is in full agreement with the previous statement. 
The fact that intermediate components of Fe II and H$\beta$ have different FWHMs ---specifically Fe II with 900 km s$^{-1}$ is greater than hydrogen broadening of 722 km s$^{-1}$--- is fully consistent with the results of  \cite{adhikari2018}. The ILR from which H$\beta$ is emitted consists of both an inward region of possibly dust-free medium and an outward region placed inside the dusty torus, as this Balmer line can be effectively produced in both of those. Therefore, the effective width seems to be lower because of higher mean Keplerian orbit. At the same time, Fe II is emitted exclusively inwards with respect to the face of the dusty torus, because dust heavily suppresses iron emission and the iron itself is depleted from the gas phase due to being trapped in grains. This results in emission from lower orbits and an effectively broader component. This may explain why the narrow component of Fe II is  so weak in the M1 B fit  \citep[see also][for appearance and disappearence of Fe II on the torus face]{He_2021}. This simple kinematic interpretation holds as the Keplerian component is predominant because of BLR and ILR coupling with the accretion disk (\cite{czerny2011} and references therein).

A global trend in the light curve is visible on timescales greater than 10 years. 
We see a decades-long steady decline in luminosity in the FUV, NUV, and optical bands with small-amplitude variations over a much shorter timescale of months in the CRTS light curve. This can be interpreted as a decrease in accretion rate, which would make the disk luminosity gradually lower. Lower ionizing flux would allow BLR to move inward \citep[if it was pushed out by strong radiation, as is the case in highly accreting AGNs; see e.g.,][]{naddaf2021}, which would cause the broad component to increase its width and possibly change its shift, which is consistent with our analysis. On longer timescales, emitting regions from higher radii should react to the change by a change in the intermediate components, with the torus face shifting inward and suppressing narrower iron emission \citep[compared to the radial emissivity study or variability analysis by][]{adhikari2018,He_2021}.
 Placing HE 0435 in the evolutionary context of a merger stage towards ULIRG and a quasar~\citep{sanders1988} is therefore interesting, although the timescale of the reported flux changes is very short in comparison to the merging episode of the hierarchical SMBH growth model.

Parameters of the HE 0435-5304 as fitted with CIGALE give an interesting overview of the host. The derived stellar mass of log$M_{star}$ [$M_{\odot}$] = 10.12 $\pm$ 0.62 at a redshift of about 0.4 places the source well inside the host regime of the AGNs~\citep{yang2018}. Taking into account SMBH mass (log$M_{BH}$ [$M_{\odot}$] = 8.3 $\pm$ 0.2) in comparison to $M_*$, the source lies within the typical population of AGNs, such as those in the samples presented by~\cite{haring2004, ding2020}. However, as presented in Fig.~\ref{fig:mstarmbh}, in a similar way to \cite{yang2018}, HE 0435-5304 is above $M_*$--$M_{BH}$ relations for active galaxy populations;
it hosts a rather massive SMBH in comparison to its stellar mass. This could possibly be explained by the fact that it undergoes a merger, which is not a stationary phase of rapid growth, while for example \cite{yang2018} assumed an accretion mode in their study. We speculate that maybe SMBH mass derivation is already affected by the new mass of a merging SMBH binary, while the stellar mass has not yet settled down to its final value.
If we consider the $SFR-M_*$ relation of a population of AGNs in the corresponding redshift bin, at its stellar mass our source shows a far higher rate of star formation than average Type 1 AGN (log$SFR \approx 0.5$, \citet{tomczak2016, speagle2014, suh2019}). However, the studied populations consist mostly of bright AGNs with broad lines and moderate X-ray luminosities. Does HE 0435-5304 therefore constitute an uncommon example of active SF in an AGN? Apparently not. The most striking examples are the ``cold quasars'', which are very bright and massive X-ray emitters categorized as unobscured sources but showing a very prominent cold dust signature at 250 $\mu$m, where SFR easily goes beyond 100 M$_{\odot}$ yr$^{-1}$, and in particular for log$L_{bol}$ = 45.8 corresponding to log $SFR \ge 3$ \citep{stanley2017,kirkpatrick2020}.

In summary, HE 0435-5304 is an AGN and a ULIRG, and its improved redshift is equal to $0.42788 \pm 0.00027$. The source is most probably going through a merger phase with active star formation. Its host mass is close to typical considering the error bar, although the SMBH mass is higher than the average for the given stellar mass. HE 0435-5304 hosts a SMBH in the moderate to high accretion mode, its place on the QMS is in the extreme type A population in light of its strong Fe II emission, and
it may depart from the border with the NLSy1 group within population A towards broader line sources. This may be interpreted as the BLR rebuilding in response to the preceding long-term drop in the UV-optical luminosity, possibly due to the decrease in accretion rate. 
 Decreases in radiation pressure and the ionization parameter allow BLRs to move inward, which may also be supported by the drop in the accretion rate of the disk itself as the wind zone in the accretion disk atmosphere could move inward as well.
HE 0435-5304 may have some form of freshly launched strong outflow at the inner radius of the BLR, which is our interpretation to explain the prominently shifted broadest line components in the AGN spectra \citep[as in the blue or very broad component in studies of][and others]{marziani2018}.
However, the presence of this effect is made uncertain by spectral distortion from atmospheric absorption influencing the spectra.
As an ULIRG, it may host a binary SMBH or a warped or precessing accretion disk which modulates parameters of the spectra over time. It would be interesting to follow up the source and verify our claims with higher confidence in the future.

\begin{acknowledgements}
We are very grateful to the anonymous referee, who help us significantly improve the manuscript.
We thank Wojtek Pych, Marzena Śniegowska, Bożena Czerny and Szymon Kozłowski for helpful discussions.
A part of the observations reported in this paper was obtained with the Southern African Large Telescope (SALT) under programs 2018-2-DDT-001.20190204 and 2018-2-DDT-001.20190207 (PI: Małgorzata Bankowicz). Polish participation in SALT is funded by grant No. MNiSW DIR/WK/2016/07.
This research was supported by the Polish National Science Center grants UMO-2018/30/M/ST9/00757, UMO-2016/23/N/ST9/01231 and UMO-2018/30/E/ST9/00082 and by Polish Ministry of Science and Higher Education grant DIR/WK/2018/12.

Małgorzata Bankowicz acknowledges support from the Copernicus Foundation for Polish Astronomy.

Based on observations made with ESO Telescopes at the La Silla or Paranal Observatories under programme ID(s) 094.A-0131(B) available in ESO data archive. Based on observations made by the Catalina Sky Survey, SkyMapper, Dark Energy Survey (DES) as distributed by the Astro Data Archive at NSF's NOIRLab.
This research has made use of the NASA/IPAC Extragalactic Database (NED), which is funded by the National Aeronautics and Space Administration and operated by the California Institute of Technology.
This research has made use of the VizieR catalog access tool, CDS, Strasbourg, France (DOI: 10.26093/cds/vizier). 
This publication make use of the following software:
Anaconda \citep{anaconda},
Astropy \citep{astropy},
CIGALE \citep{cigale},
DAOPHOT \citep{stetson1987,daophot},
Dynesty \citep{dynesty,higson2019,skilling2006},
ESO-MIDAS \citep{esomidas,midas},
IRAF \citep{iraf,iraf93},
Matplotlib \citep{matplotlib},
Mpdaf \citep{mpdaf},
Numpy \citep{numpy},
PyRAF \citep{pyraf},
PySALT \citep{pysalt},
Scipy \citep{scipy}.

\end{acknowledgements}

\bibliographystyle{aa} 
\bibliography{refs} 


\end{document}